\documentclass{ismdproc}

\usepackage{bm,amsmath,amssymb}
\long\def\comment#1{ }

\newcommand{\rmd}{{\rm d}}
\newcommand{\rme}{{\rm e}}
\newcommand{\eqn}[1]{Eq.~\eqref{#1}}
\newcommand{\beq}{\begin{eqnarray}}
\newcommand{\eeq}{\end{eqnarray}}

\begin{document}
%------------------------------------
\title{Multiparticle Dynamics in the LHC Era\footnote{
Theory Summary for ISMD2010, the 40th (XL) edition of the International
Symposium on Multiparticle Dynamics, Antwerp (Belgium), 21-25 September
2010.}}

\author{{\slshape Edmond Iancu} \\[1ex]
Institut de Physique Th\'{e}orique de Saclay, France and CERN, Theory
Division}

% please do not modify the following 5 lines
%\contribID{xy}  % will be entered by the editors
%\confID{yz}
%\acronym{ISMD2010}
%\doi            % will be entered by the editors

\maketitle

\begin{abstract}
Using the theory talks at ISMD2010 as a guidance, I present a personal
review of our current understanding of multiparticle interactions in QCD.
For more clarity, I separately consider hard, semi--hard, and soft
interactions, and I devote most of the space to those phenomena for which
progress has been recently made from first principles. Also, priority is
given to processes which are directly relevant for QCD studies at the
LHC, notably to forward particle production and ultrarelativistic heavy
ion collisions.
\end{abstract}

\section{Introduction}
\label{sec:intro}

The 2010 edition of the International Symposium on Multiparticle Dynamics
has been a privileged one, in several aspects. First, it has marked the
40th anniversary of this Symposium --- a respectable age which
demonstrates the maturity and the tradition of this series of meetings,
and also its capacity to continuously renew itself and adapt itself to
the evolutions, and the revolutions, which marked the field of
high-energy strong interactions over the last 40 years. Second, this
edition has given us the opportunity and joy to celebrate the 60
anniversary of our colleague Eddie de Wolf, who was one of the pioneers
of this field before becoming one of its pillars. Third, this edition has
marked the entrance of ISMD in the LHC era.  Indeed, this was the first
meeting in this series after the LHC started operating, so the
discussions at this meeting have naturally focused on the fresh LHC data
and the first lessons that can be drawn from them. In particular, the
results on the `ridge effect' in p+p collisions at $\sqrt{s}=7$~GeV by
the CMS collaboration~\cite{Khachatryan:2010gv} have been opportunely
released during this meeting, thus making ISMD2010 the first
international conference were these results have been presented and
debated, within a special session which has extended quite late in the
night.

The 40 years of existence of ISMD are also the years during which Quantum
Chromodynamics emerged and gradually asserted itself as the fundamental
theory of the strong interactions. The successive editions of ISMD have
closely followed this evolution and at the same time preserved a specific
identify via their preference for the original theme of this Symposium :
the study of the `multiparticle' (or many body) aspects of particle
production in hadronic collisions at high energy. But the precise content
of this general theme has continuously evolved, following the progress in
our conceptual understanding (notably in the framework of QCD) and the
rise in the energy of the accelerators.

Whereas for many years, the `multiparticle dynamics' was almost
exclusively associated with soft interactions among hadrons, which stay
outside the realm of perturbative QCD and thus elude calculations from
first principles, more recently this topic was extended to include
semi--hard or even hard interactions among partons, thus opening the era
of controlled calculations. This is appropriate because the modern day
accelerators --- HERA,  RHIC, the Tevatron and especially the LHC
--- give us access to the high--energy phase of QCD, which is
characterized by {\em high parton densities} and thus relatively weak
coupling, but also by important {\em collective phenomena} (at partonic
level), which call for many--body methods. Such collective phenomena,
which are most obvious in relation with the nucleus--nucleus collisions
at RHIC and the LHC, are also present in lepton--proton deep inelastic
scattering at HERA and in p+p scattering at the LHC, in the form of
parton saturation in the hadron wavefunctions, or of multiparton
interactions in the `underlying event'. Under the pressure of the
experimental results, notably at RHIC, it became clear that multiparton
phenomena are truly important at the present energies: they control the
gross features of particle production (like single--particle spectra,
rapidity distributions, multi--particle correlations) at moderate values
of the transverse energy ($E_T\sim 1\div 20$~GeV), and affect the
reconstruction of very hard jets ($E_T\gtrsim$~100~GeV) in the context of
the LHC. And they are most likely responsible for the `ridge effect'
alluded to above, as observed in both heavy ion (RHIC) and proton--proton
(LHC) collisions. Thus, whether one is interested in QCD {\em per se} or
merely as `background physics', one cannot make economy of a thorough
study of the {\em multiparton dynamics}. Fortunately, the theory has
followed, or even anticipated, the experimental evolution and new
formalisms, like the color glass condensate, have been developed from
first principles to deal with the physics of high parton densities. These
approaches and their consequences for the phenomenology have been
discussed at length at this 40th edition of ISMD, with conclusions to be
summarized below.

Another privileged playground for studying the dynamics in QCD at high
parton densities is {\em particle production at forward rapidities}. The
LHC has unprecedented capacities in that sense, due notably to the
forward detectors at CMS, ATLAS and LHCb. The first respective results
have not yet been released and our expectations for them are quite high,
in view of the promising, previous, results at HERA (e+p) and RHIC
(d+Au). Once they will become available, the LHC data should allow us to
decisively test our theoretical understanding of parton evolution in
perturbative QCD at high energy and in particular observe the so far
elusive `BFKL Pomeron' (possibly tamed by gluon saturation).

But however large the energy is, perturbative QCD cannot be the end of
the story. {\em Soft gluon interactions and confinement} are important on
large space--time separations and they introduce new correlations (e.g.
in the process of hadronisation) which get imprinted on the final
particle distribution. Soft interactions control observables like the
total and elastic cross sections or rapidity gaps in diffractive
processes, and they strongly influence the bulk features of the final
state --- albeit this influence becomes less and less important with
increasing energy. The study of soft interactions and more generally of
the non--perturbative aspects of high--energy scattering in QCD is one of
the traditional central themes of ISMD and it received a corresponding
attention at this edition as well. The discussions have been stimulated
by the first LHC data, which show significant deviations with respect to
the predictions of Monte--Carlo event generators. Such discrepancies
reflect the inability of the current event generators to properly deal
with the physics at, or beyond, the frontiers of pQCD, in particular with
multiparticle interactions. Whereas on the short time, such discrepancies
will likely be eliminated by new tunes (based on the LHC data) of the
existing MC codes, on the long term they should be an incentive towards
developing new types of event generators, which include more of our
present understanding of QCD
--- in particular, the recent progress with the physics of high parton
densities. However this poses serious challenges, to which I shall later
return. Besides, there will always be the problem of the genuinely
non--perturbative effects at soft momenta, for which it is difficult to
foresee any progress from first principles.

Another QCD system where the multiparton dynamics and collective
phenomena are undoubtedly important, is the deconfined phase of hadronic
matter at finite--temperature, the {\em quark--gluon plasma} (QGP). The
exploration of this phase started in the nucleus--nucleus collisions at
SPS and RHIC and it will continue at the LHC. One of the surprises coming
from RHIC and which seems to be confirmed by the first heavy ion data at
the LHC (which became available a couple of months after ISMD2010), is
that this plasma might be strongly coupled. This conclusion is still
under debate, for reasons that I will later explain, but it anyway raises
the question of the tools at our disposal for the study of strongly
coupled systems. Lattice QCD, which is our unique first--principles tool
in that sense, is giving us precious informations about the
thermodynamics of the QGP, but it cannot accommodate the dynamics of the
ephemeral phase produced in the intermediate stages of a heavy ion
collision. Some help in that sense has arrived from a rather unexpected
direction: string theory, or more precisely, the AdS/CFT correspondence
which relates a gauge theory (which shares some similarity with QCD) at
strong coupling to a string theory at weak coupling. This `duality' is
giving us some useful insight into the behaviour of a QGP at strong
coupling, that has been also reviewed at this meeting.

In the discussion above, I have mentioned several types of transverse
momentum scales, `soft', `semi--hard', and `hard', without being more
precise. This separation is pretty standard, but nevertheless let me
explain what I mean. By `soft' I refer to the QCD scale $\Lambda_{\rm
QCD}\sim 200$~MeV where physics is controlled by non--perturbative
phenomena like confinement. By `semi--hard' I mean a scale of the order
of the saturation momentum $Q_s$, which is the typical transverse
momentum of a gluon in the wavefunction of an energetic hadron (see
Sec.~\ref{CGC} for details). This scale grows with the energy and the
centrality of the collision, and with the atomic number (for a nucleus).
In fact, at the LHC, $Q_s$ should be reasonably hard: from 2 to 5 GeV
depending upon the total energy, the rapidity of the produced particles,
and the nature of the hadron (proton or lead nucleus). For processes
taking place at this scale, the coupling is relatively weak, so
perturbative techniques still apply, but the density of the participating
gluons is so high that collective phenomena cannot be neglected. The
proper treatment of such phenomena requires resummations of the
perturbation theory, that I shall describe later on. Finally, by `hard' I
mean transverse momenta much higher than $Q_s$ --- say, of the order of
the electroweak scale ($\sim 100$~GeV) or higher. In this regime, which
at the LHC is the most interesting one for searches of new physics, the
parton distributions are dilute, the `higher twist' (finite--density)
effects are truly negligible, and the QCD coupling is very small
($\alpha_s(M_Z)\simeq 0.1$), so the traditional perturbative calculations
apply. But one should not overlook the very recent results on `jet
quenching' in heavy ion collisions at the LHC, which suggest that medium
effects can strongly affect even such a very hard jet with $E_T\gtrsim
100$~GeV \cite{Aad:2010bu,Collaboration:2011sx}. It will be interesting
to see whether these data can be accommodated within perturbative QCD.

In what follows, I shall attempt a summary of the theory talks at
ISMD2010, by using the above separation of scales as a guideline and
following a path from `light to darkness': from what is conceptually best
understood
--- the hard sector --- to the longstanding, but still unsolved (and
always interesting) problem of soft interactions. Along the way, I will
mark a long stop by the semi--hard sector, where significant progress has
been realized in the recent years.

\section{High $p_T$ interactions: the quest for precision}
\label{High_pT}

\begin{figure}[hb]
\centerline{\includegraphics[width=0.35\textwidth]{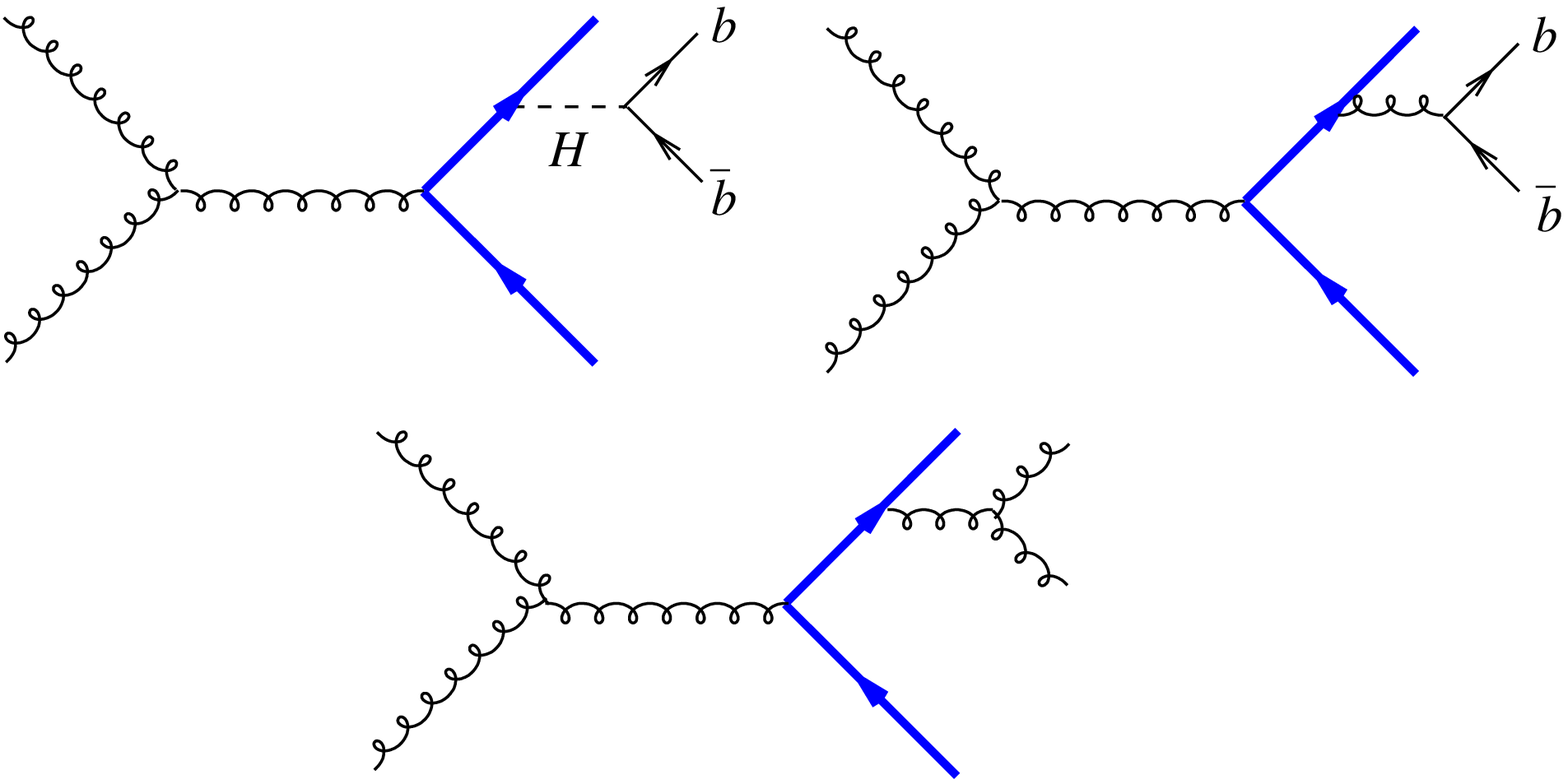}
\qquad\quad\includegraphics[width=0.45\textwidth]{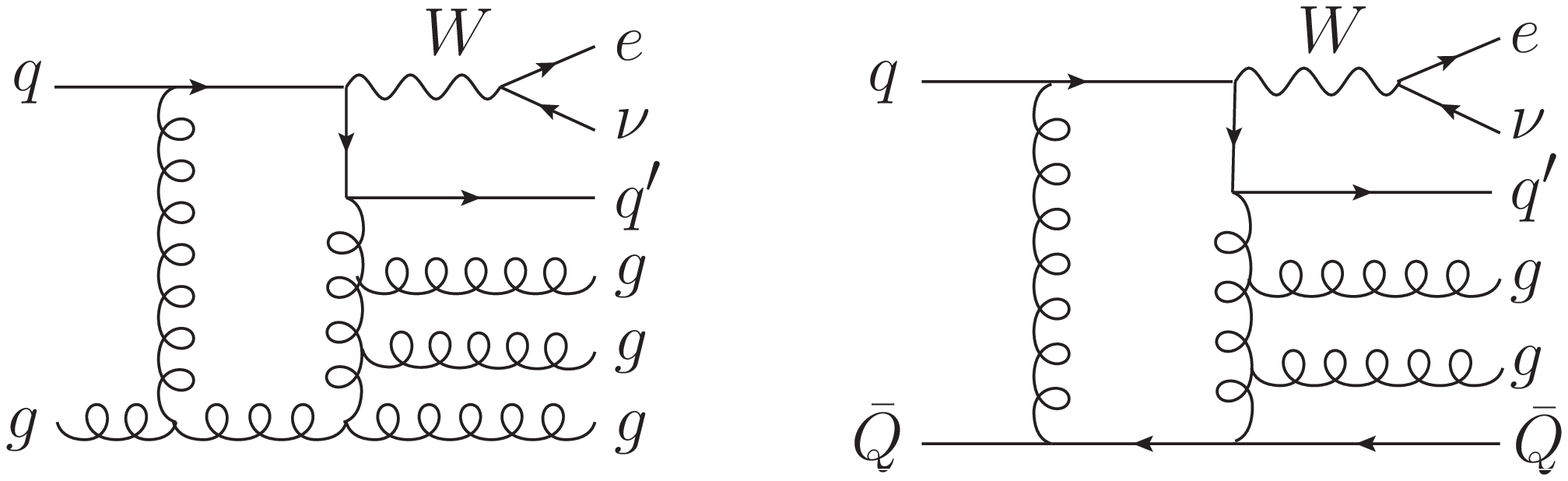}
} \caption{\sl Left: The Higgs discovery channel
$pp \to t\bar t H^*\to t\bar t b b$ together with
two background processes (shown at LO).
Right: Sample NLO diagrams for the $2\to 5$
process $pp\to W^\pm + 4$-jet.}\label{Fig:NLO}
\end{figure}

Hard processes in QCD can be accurately described within collinear
factorization, by combining partonic cross--sections computed to some
fixed order in perturbation theory --- leading--order (LO),
next--to--leading order (NLO), NNLO etc. --- with parton distribution
functions whose evolution is computed by solving DGLAP equations to the
corresponding accuracy in pQCD. (A brief review and introduction to this
session has been given by M. Grazzini~\cite{Grazzini}.) As previously
mentioned, such processes are the most interesting ones for searches of
the physics beyond the Standard Model at the LHC. For that purpose, the
corresponding rates must be known with a very good accuracy, at the NLO
level at least. Indeed, in many channels, the new physics signals could
lie in the tail of kinematic distributions and thus be hidden in broad
distributions underneath Standard Model backgrounds. For instance, the
observation of the Higgs via the production and decay chain $pp \to t\bar
t H^*\to t\bar t b b$ (a favored channel if the Higgs is relatively
light) receives backgrounds from purely QCD processes with final state
$t\bar t b b$ or even $t\bar t\, jj$, a couple of which are illustrated
in Fig.~\ref{Fig:NLO} (left). The extraction of the signal then requires
accurate predictions for the background processes, for which {\em
next--to--leading order (NLO) cross sections} in perturbative QCD are
crucial.

There are several reasons why leading--order (LO) calculations are not
accurate enough even though $\alpha_s$ is quite small. LO results depend
strongly upon the arbitrary renormalisation and factorization scales (as
introduced to define $\alpha_s$ and the parton distributions), leading to
a large uncertainty in the absolute value of the final result, which can
be reduced only by going to NLO. This problem is even sharper for
processes involving several scales like $t\bar t H$, $t\bar t + n$-jets,
$W({\rm or}\,Z) + n$-jets. Also, the shapes of distributions are first
known at NLO. Moreover, at LO a jet is modeled by the evolution of a
single parton, which is a very crude approximation; the situation can
significantly be improved by including NLO corrections.

NLO calculations involve one--loop diagrams for virtual corrections (see
the exemples in Fig.~\ref{Fig:NLO} right), whose direct evaluation
according to the familiar Feynman rules becomes prohibitively difficult
when increasing the number of external legs. To cope with that,
sophisticated methods have been developed which `automatize' the NLO
calculations, by decomposing the tensor one--loop diagrams into a basis
set of scalar one--loop diagrams with up to 4 external legs, which can be
numerically evaluated. In her talk at ISMD2010, M. Worek reported on a
recent NLO calculation for the process $pp\to t\bar t\, jj$ \cite{Worek}.
Whereas the absolute value of the NLO correction is relatively small
($\sim 10\%$), its main effect is to stabilize the result by reducing its
sensitivity to the choice of a renormalization scale, from about 35\% at
LO to $10\%$ at NLO. She has also mentioned the first ever calculation of
a $2\to 5$ process at NLO: the process $pp\to W^\pm + 4$-jet has been
recently computed (in a leading color approximation) via the unitarity
method~\cite{Berger:2010zx} (see Fig.~\ref{Fig:NLO} right).

Within the same session, R. Frederix addressed a set of three challenging
data at the Tevatron, related to the top quark phenomenology, which
exhibit a 2-$\sigma$ deviation from the respective predictions of the
Standard Model (as currently known). Such deviations could simply be
statistical fluctuations in the data, but one cannot exclude their
potential to reflect new physics BSM \cite{Frederix}. Unfortunately, the
respective processes turn out to be very sensitive to NLO, or even NNLO,
corrections (for instance, one of this observables, the forward--backward
asymmetry in the top quark pair production\footnote{Very recently, the
CDF Collaboration reported a 3.4-$\sigma$ deviation from the SM for this
quantity at large invariant mass for the $t\bar t$ pair
\cite{Aaltonen:2011kc}.}, first appears at NLO level in pQCD), as well as
to uncertainties in the matching between partonic cross--sections and
parton showers within MC event generators. Hence, one needs both more
accurate data and more accurate pQCD calculations before drawing firm
conclusions.

\section{Forward physics: the quest for BFKL and saturation}
\label{Forward}

Due to unprecedented experimental capabilities, the `forward physics',
{\em i.e.} the study of particle production at forward or backward
rapidities, very close to the collision axis, will be one of the
highlights of the experiment program at the LHC \cite{Chekanov}. This
sector too is important for new physics searches, e.g. for the Higgs
production via vector boson fusion, of via double diffractive
gluon--gluon fusion --- one of the cleanest discovery channels envisaged
so far. But at the same time this topics is very interesting for QCD {\em
per se}, in that it gives us access to the widest kinematical range for
high--energy parton evolution in QCD and thus allows us to unveil and
study phenomena like gluon saturation, multiple interactions, and the
approach towards unitarity, which are at the heart of ISMD. A better
understanding of such phenomena in the context of collider physics would
be also beneficial for the cosmic ray experiments, in improving the
modeling of the high--energy air showers: a fixed--target collision in
the air with an incoming, cosmic, particle with $E\sim 10^{17}$~eV
corresponds to p+p collisions at the LHC energy \cite{Rodriguez}.

The (pseudo)rapidity of a particle produced at an angle $\theta$ with
respect to the collision axis is $\eta=-\ln\tan(\theta/2)$, so large
rapidity (in absolute value) is tantamount to small angle, or relatively
small transverse momentum $k_\perp=E\sin\theta$ (for a given particle
energy $E$). This is why forward physics was traditionally associated
with soft particle production. However, the situation has changed with
the advent of the modern--day accelerators, HERA, RHIC and especially the
LHC, where the large center--of--mass energy makes it possible to produce
hard particles at small angles and in particular study jet physics in the
forward region.

For instance, the forward calorimeters at ATLAS and CMS can mesure jets
with $k_\perp\ge 30$~GeV up to rapidities as large as $\eta\simeq 6$.
This looks like truly `hard' physics, in the sense that the relevant
values of the QCD coupling are small, but at the same time --- due to the
high energy and the forward kinematics (see below) --- this explores the
very low--$x$ region (with $x$ denoting the longitudinal momentum
fraction of a parton) of the wavefunction of one of the incoming hadrons,
where the gluon density is quite high even for such large transverse
momenta and therefore usual perturbative techniques, like collinear
factorization or the traditional notion of a `parton distribution', fail
to apply. This is the {\em semi--hard} region, where an intrinsic
transverse momentum scale, associated with the gluon density, emerges in
the hadron wavefunction --- the saturation momentum $Q_s$ --- and
perturbation theory needs to be reorganized, even if the coupling is
weak, to account for high energy radiative corrections (BFKL evolution)
and `higher--twists' effects (gluon saturation), which now become of
order one.

\begin{figure}[thb]
\centerline{
 \includegraphics[width=0.28\textwidth]{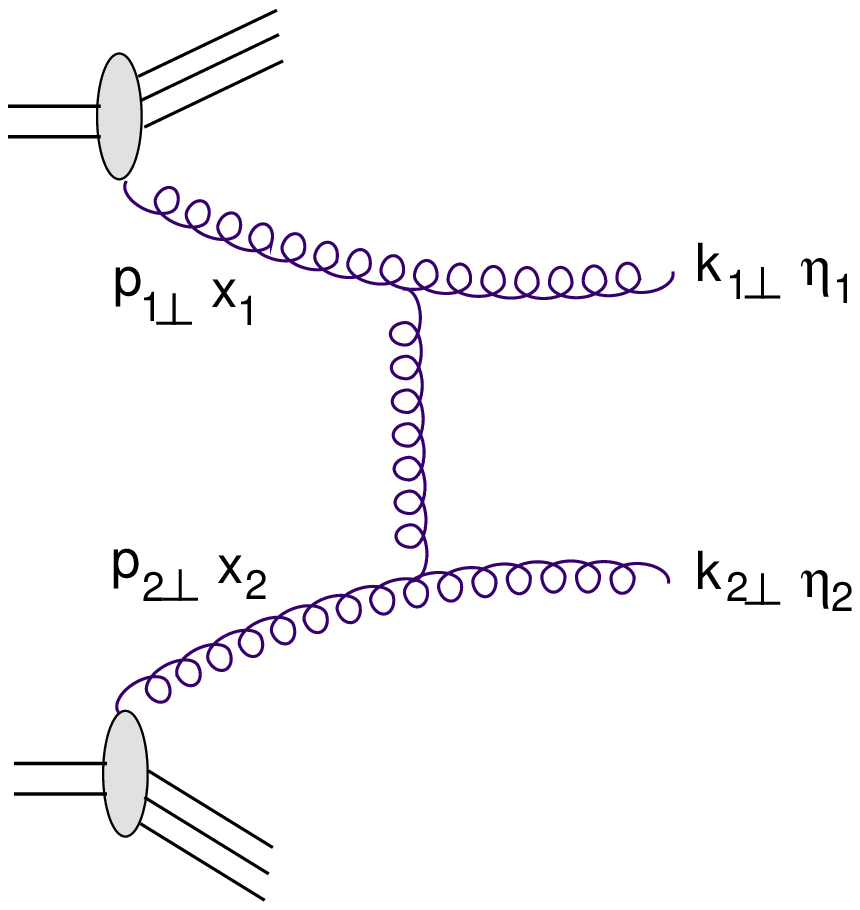}
 \includegraphics[width=0.33\textwidth]{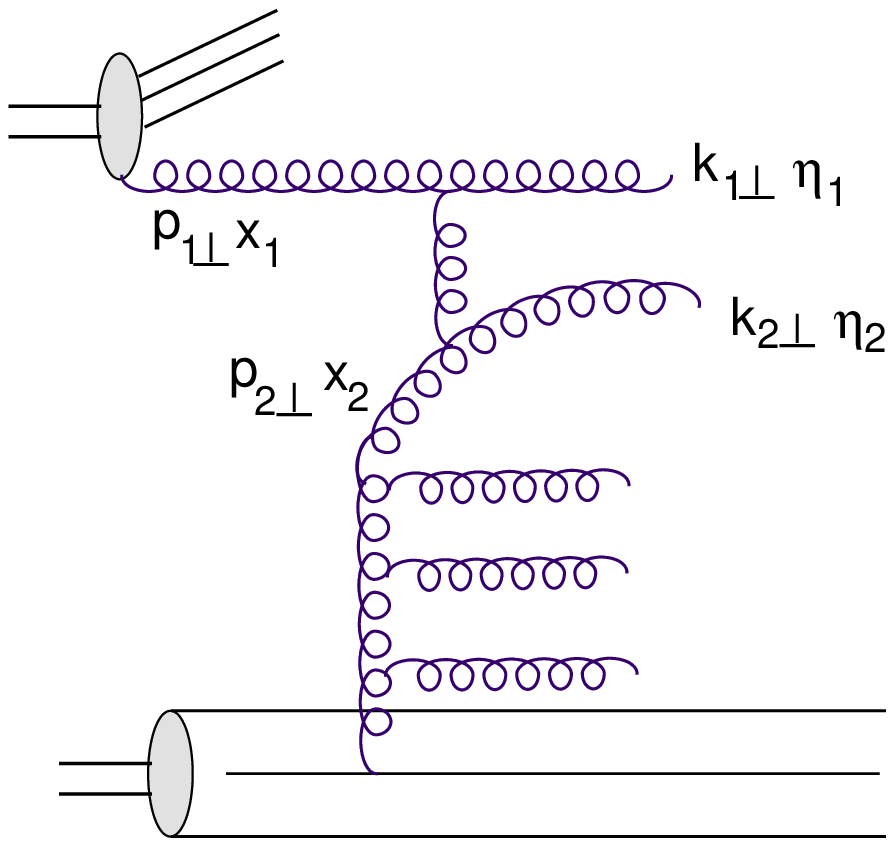}\qquad
 \includegraphics[width=0.28\textwidth]{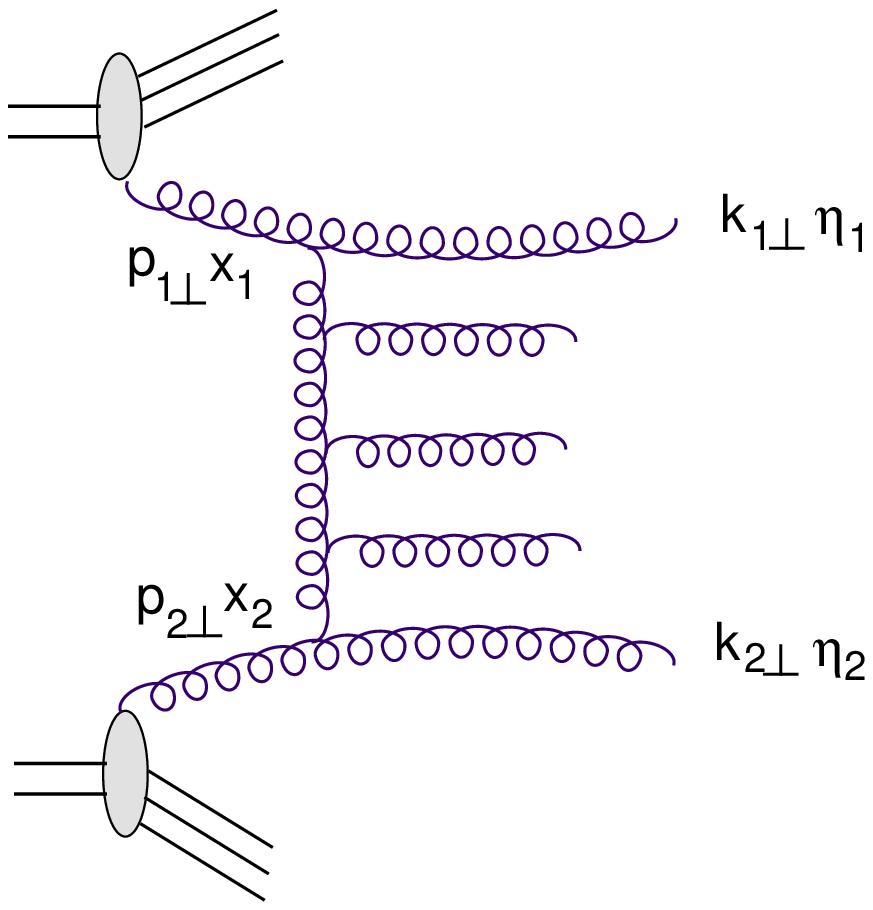}}
 \caption{\sl Left: Leading--order diagram for the production of a pair
 of jets in a proton--proton collision. Middle: When both jets are forward,
 one needs to perform a BFKL resummation of the gluon distribution in the
 `target' (lower) proton. Right: Forward--backward, or Mueller-Navelet,
 jets. Now the BFKL resummation refers to the partonic cross--section.
\label{Pair}}
\end{figure}

In order to describe the essence of these resummations in a specific
physical context, let me schematically address the problem of the
production of a pair of jets in a hard scattering in hadron--hadron
collisions (say, at the LHC). A diagram contributing to this process to
LO in pQCD is displayed in Fig.~\ref{Pair} left. Within collinear
factorization, one assumes that the two partons which scatter with each
other have negligible transverse momenta: $p_{1\perp} = p_{2\perp} =0$.
Then transverse momentum conservation requires that ${\bm k}_{1\perp}
=-{\bm k}_{2\perp}$, whereas the conservation of the energy and the
longitudinal momentum determines the longitudinal momentum fractions
$x_1$ and $x_2$ of the incoming partons as a function of the jets
kinematic variables --- their (equal) transverse momenta  $|{\bm
k}_{1\perp}| = |{\bm k}_{2\perp}| \equiv k_\perp$ and the
(pseudo)rapidities $\eta_1$ and $\eta_2$ :
\begin{eqnarray}
 \comment{ x_1\,=\,\frac{k_{\perp}}{\sqrt{s}}\,\Big[{\rm e}^{\eta_1}
+{\rm e}^{\eta_2}\Big]\,,\qquad x_2
\,=\,\frac{k_{\perp}}{\sqrt{s}}\,\Big[{\rm e}^{-\eta_1}+ {\rm
e}^{-\eta_2}\Big].}
x_1\,=\,\frac{k_{1\perp}}{\sqrt{s}}\,\rme^{\eta_1}
+\frac{k_{2\perp}}{\sqrt{s}}\,\rme^{\eta_2}\,,\qquad
x_2 \,=\,\frac{k_{1\perp}}{\sqrt{s}}\,\rme^{-\eta_1}+
\frac{k_{2\perp}}{\sqrt{s}}\,\rme^{-\eta_2}\,.
  \end{eqnarray}
This yields the following estimate for the cross--section for di-jet
production:
 \begin{eqnarray}\label{coll}
  \frac{\rmd \sigma}{\rmd^2 k_{1\perp}
\rmd^2 k_{2\perp}\rmd\eta_1 \rmd \eta_2}=\sum_{ij}
x_1 f_{i}(x_1,\mu^2)\, x_2 f_{j}(x_2,\mu^2)\,
\delta^{(2)}({\bm k}_{1\perp}+{\bm k}_{2\perp})
\frac{\rmd \hat\sigma_{ij}}{\rmd k^2_{\perp}}\,,
 \end{eqnarray}
with ${\rmd \hat\sigma}/{\rmd k^2_{\perp}}\propto \alpha_s^2/k^4_{\perp}$
at high energy. Under these assumptions, a measurement of the azimuthal
correlation ${\rm d}\sigma/{\rm d}\Delta\phi$, with $\Delta\phi$ the
angle between ${\bm k}_{1\perp}$ and ${\bm k}_{2\perp}$, would yield a
sharp peak at $\Delta\phi=\pi$. This peak will get somewhat smeared after
the inclusion of NLO corrections to the hard matrix element, but it will
always be quite sharp within the context of collinear factorization,
because the probability for emitting additional, hard, partons is small
at weak coupling. However, this cannot be the physical reality at forward
kinematics, as I explain now.

To be specific, I shall choose the situation where both $\eta_1$ and
$\eta_2$ are positive and large, and relatively close to each other:
$\eta_{1,2}\gtrsim 4\div 5$, with $|\eta_1-\eta_2|\lesssim 1$. This
corresponds to the production of a pair of `forward di-jets', a process
that has been already measured in d+Au collisions at RHIC, with some
interesting results \cite{Braidot:2010zh,Marquet,Albacete} that I shall
later discuss. In this case, we have $x_1\gg x_2$, so this process probes
{\em very asymmetric parton configurations}. For p+p collisions with
$\sqrt{s}=7$~TeV and $k_\perp=35$~GeV, we have
%${k_{\perp}}/{\sqrt{s}}=5\times 10^{-3}$, and then
$x_1\gtrsim 0.2$ and $x_2$ between $3\times 10^{-5}$ and $10^{-4}$.
% when
%$\eta_{1,2}\sim 4$, or even $x_2\simeq 10^{-5}$ when $\eta_{1,2}\sim 6$.
Thus, remarkably, the forward kinematics at the LHC allows us to probe
very small values of $x$ while staying in the pQCD--controlled regime at
hard transverse momenta (unlike what happened at HERA, where small--$x$
was accompanied by low $Q^2$). For such small values of $x_2$, there is a
large rapidity interval $Y_2\equiv \ln(1/x_2)\sim 10$ for {\em
high--energy evolution} via gluon emission in the `target' proton (the
proton which moves oppositely to the two jets). Indeed, the differential
probability for the emission of a `soft' ($x\ll 1$) gluon from some other
parton via bremsstrahlung reads ($C_R$ is a color factor)
  \begin{eqnarray}\label{brem} \rmd P_{\rm Brem}\,\simeq\,\frac{\alpha_s
 C_R}{2\pi^2}\,\frac{\rmd^2k_\perp}{k_\perp^2}\,\frac{\rmd
 x}{x}\,,\end{eqnarray}
showing that there is a probability of order $\alpha_s$ to emit a gluon
per unit rapidity $Y=\ln(1/x)$. When $\alpha_s Y \gtrsim 1$, we have a
large phase--space for soft gluon emission, leading to a rapid increase
in the gluon distribution at small $x$, via gluon cascades as that
illustrated in Fig.~\ref{Pair} middle. Such a growth, which is clearly
seen in the DIS results at HERA \cite{Amanda}, can also be mimicked (at
least over a limited interval in $y$) by the DGLAP evolution of the
PDF's; but this is not the proper way to describe the gluon evolution
with decreasing $x$, since there is no transverse momentum ordering in
the successive emissions of soft gluons. Rather the transverse kinematics
must be dealt with exactly, and this is what is done in the BFKL
(Balitsky-Fadin-Kuraev-Lipatov) equation \cite{IancuBFKL}, which resums
the radiative corrections of order $(\alpha_s\ln 1/x)^n$ for any $n$.
This introduces some formal complications: one needs to use {\em
unintegrated} parton distributions (uPDF), which also keep trace of the
parton transverse momentum $k_\perp$ (in addition to its longitudinal
fraction $x$), together with an appropriate factorization scheme, `the
$k_T$--factorization', which is known to hold to leading logarithmic
accuracy (LLA) at high energy \cite{Hautmann}.

Specifically, the $k_T$--factorized version of \eqn{coll} reads
\begin{eqnarray}\label{kT}
\frac{\rmd \sigma}{\rmd^2 k_{1\perp}
\rmd^2 k_{2\perp}\rmd\eta_1 \rmd \eta_2}&\!=\!&\sum_{ij}\!
\int\rmd^2 {\bm p}_{1\perp}\rmd^2 {\bm p}_{2\perp}
\,\delta^{(2)}({\bm p}_{2\perp}+{\bm
p}_{1\perp}-{\bm k}_{1\perp}-{\bm k}_{2\perp})\nonumber\\
&{}&\qquad\quad\times\
\Phi_i({\bm p}_{1\perp},x_1)\,
\frac{\rmd \hat\sigma_{ij}}{\rmd \hat t}\,
\Phi_j({\bm p}_{2\perp},x_2)
 \,,\end{eqnarray}
where $\hat t=({\bm k}_{1\perp}-{\bm p}_{1\perp})^2$ and
 $\Phi_i({\bm p}_{\perp},x)$ represents the unintegrated
parton distribution (uPDF) for parton species $i$, that is, the number of
partons per unit transverse momentum per unit rapidity. The standard,
`integrated', PDF $x f_{i}(x,\mu^2)$ is obtained by integrating
$\Phi_i({\bm p}_{\perp},x)$ over ${p}_{\perp}$ up to the factorization
scale $\mu$. In the context of the high--energy resummations, the use of
uPDF's is restricted to gluons, since the respective distribution is the
only one to be amplified with decreasing $x$. However, as recalled by
I.~Cherednikov~\cite{Cherednikov} at ISMD2010, uPDF's appear also in
other contexts (often under the name of `transverse momentum dependent
parton densities', or TMD's), like the study of spin asymmetries in
semi--inclusive DIS, or the $k_\perp$ distribution in the Drell--Yan
process. At a formal level, they are defined as phase--space
distributions (or Wigner functions), but in general this formal
definition meets with ambiguities associated with overlapping ultraviolet
divergences, or with the gauge--links required by gauge
invariance~\cite{Cherednikov}. Such difficulties disappear in the context
of high--energy scattering (at least to the LLA of the
$k_T$--factorization), where the would--be UV divergences associated with
the rapidity are taken care by the BFKL evolution and the gauge--links
are unambiguously identified as eikonal Wilson lines, to be discussed
below.

The collinear factorization in \eqn{coll} can be formally recovered from
\eqn{kT} by neglecting the sum ${\bm p}_{2\perp}+{\bm p}_{1\perp}$ of the
parton transversa momenta in the $\delta$--function inside the integrand.
However, such an approximation would become incorrect for sufficiently
small values of $x$ : BFKL equation predicts not only a rapid rise in the
number of gluons with decreasing $x$, but also the fact that the typical
transverse momenta of these gluons can deviate quite strongly from the
starting value at large $x$
--- the more so, the smaller $x$ is. This should not be a surprise: in
the absence of any ordering in $k_{\perp}$, the evolution proceeds as a
random walk in the transverse momentum space, with the rapidity $Y=\ln
1/x$ playing the role of an `evolution time'.

The most salient features of the BFKL evolution can be appreciated on the
basis of its following, simplified, form:
\beq\label{BFKL}
 \frac{\partial \Phi_g(\rho,Y)}{\partial Y}\,\simeq\,\omega\alpha_s
 \Phi_g
 \,+\,\chi\alpha_s \partial_\rho^2 \Phi_g\,,
 \eeq
where $\omega$ and $\chi$ are numerical constants (the `BFKL intercept'
and `diffusion coefficient') and $\rho=\ln(k_{\perp}^2/Q_0^2)$, with
$Q_0$ the typical transverse momentum at the rapidity $Y_0$ at which one
starts the evolution. The first term in the r.h.s. describes gluon
multiplication and predicts an exponential increase with $Y$ : $\Phi_g
\propto \exp(\omega\alpha_s Y)$. The second term describes diffusion in
$\rho$ and predicts that the typical transverse momenta of the emitted
gluons can deviate from $Q_0$ according to $|\ln(k_{\perp}^2/Q_0^2)|
\lesssim \sqrt{\chi\alpha_s Y}$. Note that this diffusion in symmetric in
$\rho$ : the transverse momenta of the emitted gluons can be either
harder or softer than the original $Q_0$. If extrapolated to very large
values of $y$, these features become problematic: the rapid rise of the
gluon distribution eventually leads to violations of the unitarity bounds
for the scattering amplitudes; and the BFKL diffusion eventually enters
the non--perturbative regime at soft momenta $k_{\perp}\sim\Lambda_{\rm
QCD}$, where this whole approach becomes unreliable. I shall later argue
that such problems are solved by {\em gluon saturation} within a
consistent approach at weak coupling.

%get postponed after adding NLO corrections and are eventually

In particular, for the asymmetric situation $x_1\gg x_2$ corresponding to
the production of a pair of forward di-jets, the high--energy evolution
is important only for the `target' proton, for which $Y_2\sim 10$. As for
the `projectile' proton (for which $Y_1\sim 1$), this can still be
treated in the spirit of the collinear factorization. By neglecting ${\bm
p}_{1\perp}$ next to ${\bm p}_{2\perp}$ in \eqn{kT}, using the
$\delta$--function to integrate over ${\bm p}_{2\perp}$, and keeping only
the gluon distribution in the target (since this is the only one to be
enhanced at small $x_2$) one finds
\begin{eqnarray}\label{asym}
\frac{\rmd \sigma}{\rmd^2 k_{1\perp}
\rmd^2 k_{2\perp}\rmd\eta_1 \rmd \eta_2}
 &\simeq& \sum_{i}\, x_1 f_{i}(x_1,\mu^2)\, \frac{\alpha_s^2}
 {k_{1\perp}^4}\,\Phi_g({\bm k}_{1\perp}+{\bm k}_{2\perp},
 x_2)
 \,.\end{eqnarray}
This formula makes it clear that it is the BFKL evolution of the gluon
distribution in the target which controls the energy and rapidity
dependencies of the di-jet cross-section, as well the transverse momentum
unbalance (${\bm k}_{1\perp}+{\bm k}_{2\perp}\ne 0$) between the produced
jets. Namely, the cross--section grows very fast with the COM energy, as
the power $s^{\omega\alpha_s/2}$. For a fixed COM energy, it rises
exponentially with the rapidities $\eta_1\simeq\eta_2$ of the produced
jets. Also, the distribution in ${\bm k}_{1\perp}+{\bm k}_{2\perp}$
broadens rapidly with increasing $\eta_i$, due to the BFKL diffusion.
Within the present approximations, the value of ${\bm k}_{1\perp}+{\bm
k}_{2\perp}$ in a given event is equal to minus the sum of the transverse
momenta of the gluons radiated within the BFKL cascade (cf.
Fig.~\ref{Pair} middle) and which are liberated in the final state. This
also shows that the structure of the event is modified by the high energy
evolution of the hadron wavefunctions.

\eqn{asym} is based on the LLA at high energy, which is rather poor: not
only the NLO corrections to the BFKL equation are known to be large
\cite{IancuNLBFKL,IancuSalam99}, but there are also important non--linear
effects (gluon saturation) associated with the high gluon density in the
target \cite{Gelis:2010nm}. But before turning to a discussion of such
effects, let me explain why the process described so far --- the
production of a pair of forward jets --- is not necessarily optimal for a
study of high--energy evolution at the LHC (although it was very useful
in that sense at RHIC, as I shall later review). This discussion will
also allow me to introduce a process which is better suited for the
kinematics at the LHC: the Mueller--Navelet jets \cite{Mueller:1986ey}.

Specifically, the process in \eqn{asym} involves two hard transverse
momentum scales, ${k}_{1\perp}$ and ${k}_{2\perp}$, which in the context
of the LHC cannot be smaller that 20~GeV (for the produced jets to be
distinguishable from the hadronic background). These scales must be
compared to the maximal transverse momentum that can be generated in the
target proton wavefunction via the BFKL diffusion. More precisely, I
shall shortly argue that the relevant comparison scale is the {\em target
saturation momentum}, which grows with $1/x$ and for a proton at $x\sim
10^{-5}$ is expected (from analyses of the HERA data) to be $Q_s(x)\sim
1$~GeV \cite{Albacete}. This value is considerably smaller than the
${k}_{\perp}$ of the external jets, meaning that, first, the asymmetry
expected on the basis of the BFKL evolution is rather tiny and thus
difficult to observe and, second, the process also involves large
collinear logarithms $\ln({k}_{\perp}^2/Q_s^2)$ which need to be resummed
in the parton distributions, on top of the high--energy logarithms
$\ln(1/x_2)$.

The simultaneous presence of several, large, scales calling for different
types of resummations is a generic feature of the forward  physics, which
often complicates the corresponding theoretical analysis. Yet, there
exists a process for which this problem is less pronounced and which can
create a larger asymmetry between the produced jets via the high--energy
evolution. This is the production of a pair of forward--backward, or
Mueller--Navelet (MN), jets \cite{Mueller:1986ey}. In this case $\eta_1$
and $\eta_2$ are large but of opposite signs, say $\eta_1 >0$ and $\eta_2
< 0$, so that both $x_1$ and $x_2$ are relatively large, $\sim 0.1$, and
the collinear factorization in \eqn{coll} is perfectly legitimate. In
particular, the large collinear logarithms $\ln({k}_{\perp}^2/Q_0^2)$ are
absorbed in the standard way, in the PDFs. What is peculiar about this
process is the large rapidity separation {\em between} the two jets,
$Y\equiv\eta_1-\eta_2$ with $Y\simeq 10$ at the LHC, which favors the
BFKL evolution of the {\em partonic} cross--section $\hat\sigma$  (see
Fig.~\ref{Pair} right). Note however that this evolution now refers to
partons which are {\em hard to start with}, and not to protons.
Accordingly, the typical momenta of the BFKL gluons are comparable to
those of the final jets, thus permitting a {\em large asymmetry} between
the measured momenta ${k}_{1\perp}$ and ${k}_{2\perp}$. But these final
momenta are still {\em comparable} with each other, so there is no large
collinear log $\ln({k}_{1\perp}^2/{k}_{2\perp}^2)$ to worry about. Thus,
this process offers a clean set--up to test the BFKL evolution and is
indeed under intense scrutinize at the LHC.

But in order to be conceptually meaningful and practically useful, the
BFKL calculations must be extended to include NLO corrections and
saturation effects. I have previously mentioned that the NLO corrections
to the BFKL equation, {\em i.e.} the effects of order $\alpha_s(\alpha_s
\ln 1/x)^n$ in perturbation theory, turn out to be very large
\cite{IancuNLBFKL} and in fact some effort was needed to render them
meaningful in practice. This required a better understanding of the
interplay between the high--energy radiative corrections and the
collinear ones, leading to `RG--improved' evolution equations which
partially resums both types of effects \cite{IancuSalam99}. These
equations predict that the high--energy evolution is considerably slower
than predicted by the LO BFKL equation: both the rise of the gluon
distribution with $1/x$ and the transverse momentum diffusion are
drastically reduced by the NLO corrections, but they are not completely
washed out. (For instance, the NLO BFKL calculation of MN jets in
\cite{Colferai:2010wu} leads to results which are quite close to the
respective predictions of the NLO DGLAP formalism.) In particular, the
conceptual problems of the BFKL evolution are not cured by NLO
corrections, but merely postponed to higher energies. Such energies, at
which unitarity corrections become important even for relatively hard
momenta, have been already reached by the modern day accelerators, as
clearly shown by the need to include multiparticle interactions and
`infrared cutoffs' which rise with the energy (and thus mimic saturation;
see below) in any Monte--Carlo code aiming at describing the exclusive
final state at RHIC or the LHC.

\begin{figure} [hb]
\centerline{
 \includegraphics[width=0.35\textwidth]{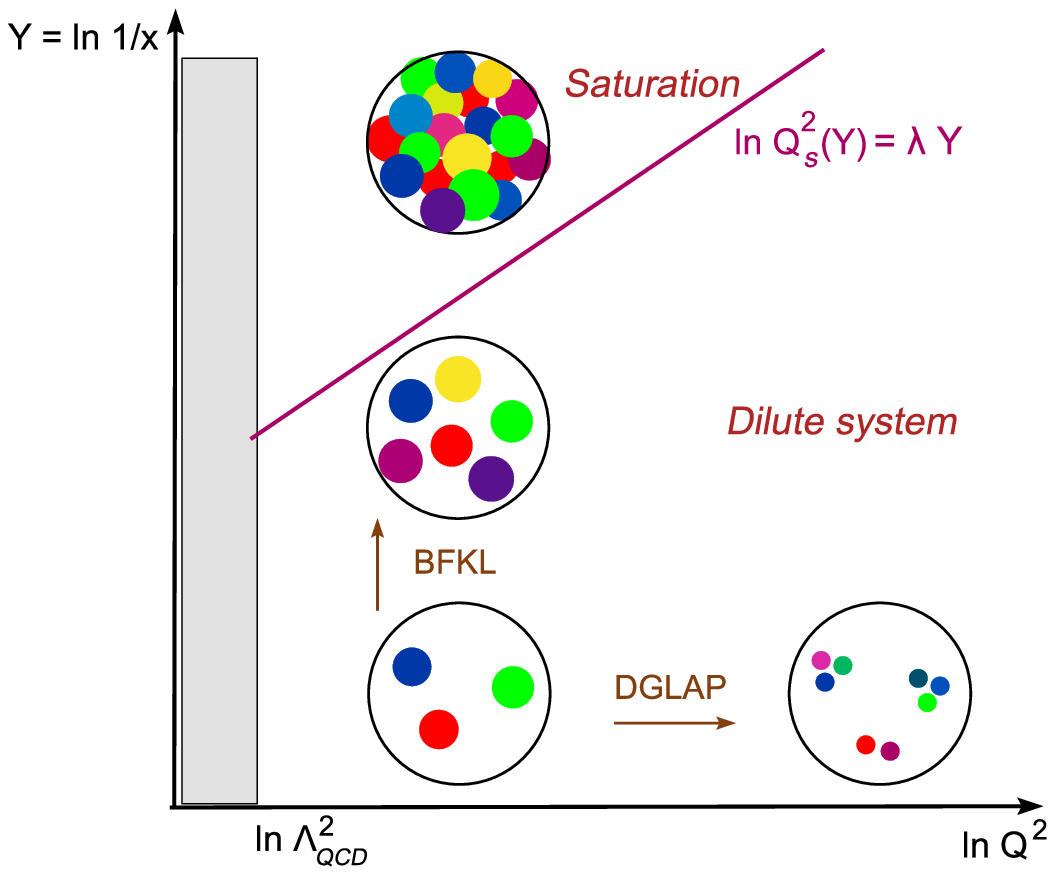}\qquad\
 \includegraphics[width=0.55\textwidth]{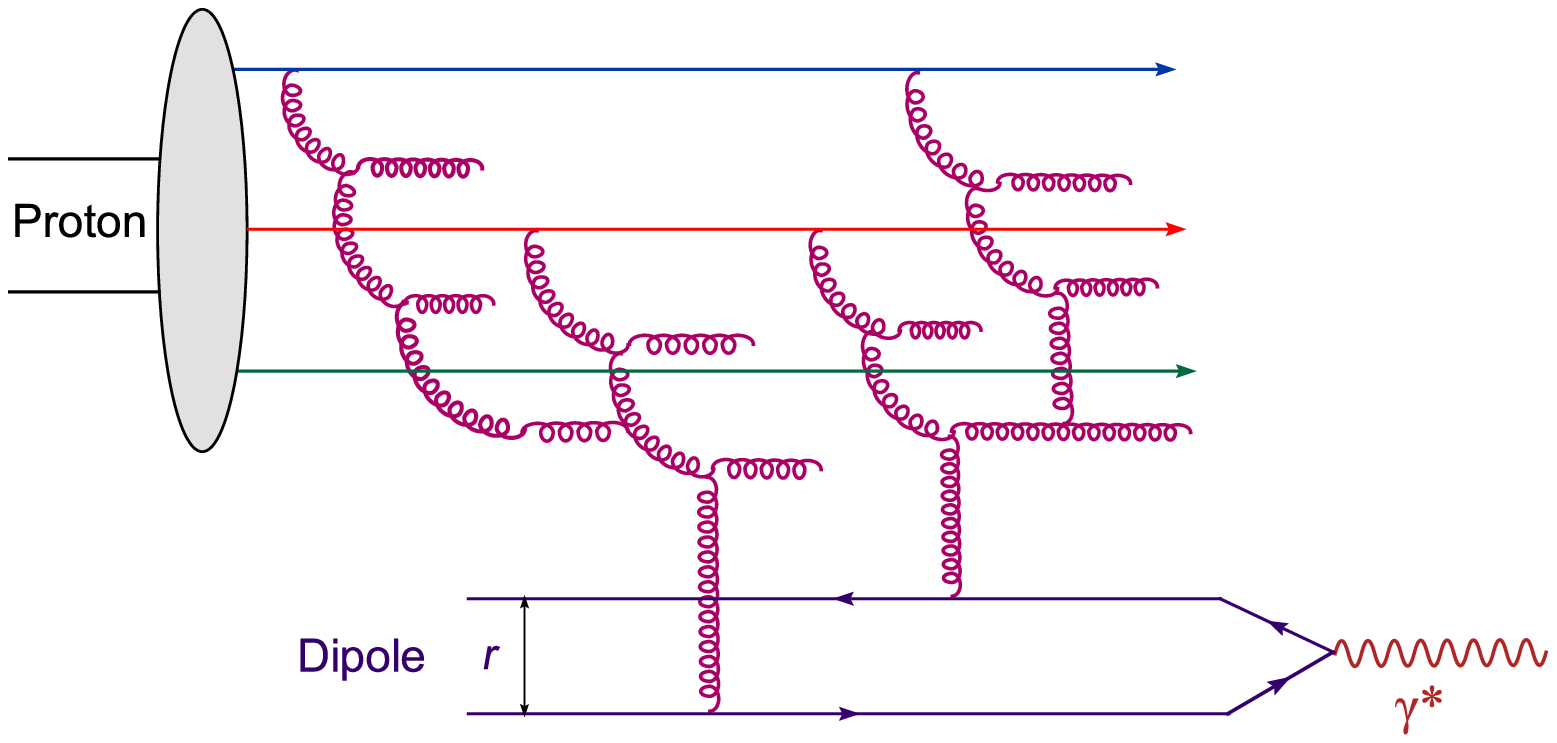}}
 %\vspace*{-.6cm}
\caption{\sl Left: A
`phase--diagram' for parton evolution in QCD; each
colored blob represents a parton with transverse area $\Delta x_\perp
\sim 1/Q^2$ and longitudinal momentum $k_z=xP_z$.
Right: Gluon evolution in the presence of saturation, as probed
via multiple scattering in DIS.
\label{Fig_CGC}}
\end{figure}

A consistent description of such phenomena from first principles requires
in particular the understanding of non--linear phenomena in parton
evolution at small $x$ together with new factorization schemes which go
beyond the single--particle PDFs in such a way to capture the
correlations associated with these non--linear phenomena. This is the
topics of {\em gluon saturation} which met with important progress over
the last years \cite{Gelis:2010nm} and represented a main focus for two
of the sessions of ISMD2010 --- `Forward physics' and `High--density QCD'
(see Sec.~\ref{CGC} below). To concisely describe this topics, let me
refer to the cartoon of parton evolution depicted in Fig.~\ref{Fig_CGC}
left. As shown in this figure, there are two main directions of evolution
--- the DGLAP evolution with increasing virtuality $Q^2$ (or transverse
momentum $k_\perp^2$) and the BFKL evolution with increasing energy, or
rapidity $Y=\ln 1/x$ ---, which both lead to a rise in the number of
partons, via parton branching, but with very different consequences.

By the uncertainty principle, a parton has a transverse area $1/Q^2$, so
with increasing $Q^2$ the total area occupied by the partons decreases
much faster than the rise $\propto\ln Q^2$ in the number of partons;
accordingly, the DGLAP evolution leads to a partonic system which is more
and more {\em dilute}. By contrast, when increasing $Y$, the transverse
momenta vary only slowly and symmetrically (via BFKL diffusion), so the
gluons emitted within the BFKL evolution are roughly of the same size,
whereas their number grows exponentially with $Y$. So, clearly, this
evolution produces a {\em dense} system of partons which overlap with
each other. At high density, the coupling is weak, so the gluon
interactions are suppressed by $\alpha_s\ll 1$, but they are enhanced by
the gluon {\em occupation number} $n(k_{\perp},Y)$, since a gluon can
interact with all the other gluons that it overlaps with. Here $n$ is
roughly the unintegrated gluon distribution per unit transverse area,
$n\simeq \rmd\Phi_g/\rmd^2 b_\perp$, and it grows very fast, $n\propto
\exp(\omega\alpha_s Y)$, within the BFKL approximation. Thus, despite of
the weakness of the coupling, the system becomes fully non--linear when
$Y$ is large enough for $n$ to become of order $1/\alpha_s$. When this
happens, the gluons start to repeal each other, which inhibits the
emission of new gluons, thus taming the BFKL rise of the gluon
distribution. This is `gluon saturation'.

The theoretical description of this phenomenon within perturbative QCD is
extremely complex, since the non--linear effects generate many--body
correlations whose high--energy evolution is entangled. After pioneering
work in the mid eighties \cite{IancuGLR}, the pQCD description of this
non--linear evolution has been constructed over the last 15 years, in the
form of an effective theory for the small--$x$ gluons in the wavefunction
of an energetic hadron: the {\em color glass condensate} (CGC)
\cite{Gelis:2010nm}. The `condensate' refers to the strong, classical,
color field describing a gluon configuration with large occupation
numbers, while the `glass' reminds that this configuration is randomly
generated via the high--energy evolution. Thus, to compute observables
one needs to average over all the classical field configurations, with
the `CGC weight function' (a functional encoding the multi--gluon
correlations at small $x$). The central equation of this effective theory
is a functional renormalization group equation, known as `JIMWLK' (from
Jalilian-Marian, Iancu, McLerran, Weigert, Leonidov, and Kovner), which
shows how the CGC weight function gets built via soft gluon emission in
the background of a strong color field. This equation has been
constructed in pQCD, to leading--log accuracy in the radiative
corrections at high energy, but to all orders in the `higher--twist'
effects associated with the high gluon density. In the multi--color limit
$N_c\gg 1$, it reduces to a single, non--linear, equation for the gluon
occupation number --- a non--linear generalization of the BFKL equation
known as the Balitsky--Kovchegov (BK) equation, which has recently been
extended to NLO accuracy \cite{IancuRunBK,Balitsky:2008zza}.

A cartoon version of BK equation, to be compared to the `BFKL equation'
\eqref{BFKL}, reads
 \beq\label{BK}
 \frac{\partial n(\rho,Y)}{\partial Y}\,\simeq\,\omega\alpha_s
 n
 \,+\,\chi\alpha_s \partial_\rho^2 n-\alpha_s^2 n^2\,
 \eeq
As compared to \eqn{BFKL}, this involves an additional, quadratic, term
with a negative sign, which describes gluon recombination: when $n\sim
1/\alpha_s$, the linear and non--linear terms in the r.h.s. cancel each
other and then the gluon distribution stops rising.  Due to the
non--locality of this equation in $\rho\equiv\ln(k_{\perp}^2/Q_0^2)$, the
saturation domain progresses with $Y$, towards larger and larger values
of $k_\perp$. This progression is characterized by the {\em saturation
momentum} $Q_s(Y)$ --- the value of $k_{\perp}$ below which one finds
saturation at a given $Y$. Namely, the BK equation implies
 % \beq \label{Qs}
 %Q_s^2(Y)\,\simeq\,Q_0^2 \,\rme^{\lambda_s(Y-Y_0)}\,,\eeq
\beq\label{dndy}
 n(k_{\perp},Y)\,\approx
 \begin{cases}
        \displaystyle{\frac{1}{\alpha_s}\,} &
        \text{ if\ $k_\perp\lesssim Q_s(Y)$, }
        \\*[0.35cm]
        \displaystyle{\frac{1}{\alpha_s}\left(
        \frac{k_\perp^2}{Q_s^2}\right)^{\gamma_s}} &
        \text{ if\ $k_\perp \gg Q_s(Y)$,}
        \end{cases}
  \eeq
where $\gamma_s\simeq 0.63$ and $Q_s^2(Y)\simeq Q_0^2
\,\rme^{\lambda_s(Y-Y_0)}$ with $\lambda_s\simeq 0.3$ at NLO accuracy
\cite{Triantafyllopoulos:2002nz}.

The initial conditions $Y_0$ and $Q_0$ for the high--energy evolution
depend upon the target. For a hadronic target, one generally takes
$Y_0\simeq 4$ (or $x_0\simeq 10^{-2}$), which ensures that $\alpha_s
Y_0\sim 1$. The corresponding value of $Q_0$ is non--perturbative and it
is fitted from the data. The fits to the DIS structure function at HERA
using BK equation with running coupling yield $Q_0^2\simeq 0.2$~GeV$^2$
for a proton at $x_0=10^{-2}$ \cite{Albacete}. For a heavy nucleus with
atomic number $A\gg 1$ and for central collisions, we expect $Q_0^2$ to
be enhanced by a factor $A^{1/3}$ w.r.t. a nucleon, because of the
correspondingly larger transverse density of valence quarks available for
initiating the evolution \cite{Gelis:2010nm}. When the evolution starts
directly with a hard parton, so like for MN jets, $Q_0$ is the (hard)
transverse momentum of that parton; but this becomes a `saturation scale'
only for a sufficiently large $Y_0$, such that the perturbative evolution
of the original parton up to $Y_0$ be able to build a relatively dense
system of gluons. This value $Y_0$ can be computed in pQCD, which
predicts $Y_0\simeq 8\div 10$ independently of $Q_0$ \cite{Iancu:2008kb}.
These estimates imply that the saturation momenta to be explored in the
forward kinematics ($|\eta|\gtrsim 5$) at the LHC can be as large
$Q_s^2\simeq 2$~GeV$^2$ for protons, $Q_s^2\simeq 6$~GeV$^2$ for lead
nuclei, and $Q_s^2\simeq 100$~GeV$^2$ for MN jets with $Y\gtrsim 10$. The
last number is particularly striking: it shows that by focusing on the
high--energy evolution a `hot spot' inside a hadron (a parton with high
$Q^2$ and large $x$), one can reach values of the saturation momentum
which are much higher than the {\em average} saturation momentum in that
hadron. But this comes with a price: a parton with large $x\sim 0.1$ and
high $Q^2$ represents a rare fluctuation in the proton evolution, hence
the corresponding PDF $f(x,Q^2)$ is very small.

\eqn{dndy} has some important consequences:

\texttt{(i)} It shows that the typical transverse momenta in the
wavefunction of an energetic hadron are of order $Q_s$ and thus they
become hard for sufficiently large $Y$. Hence, a weak--coupling approach
is indeed appropriate for the study of the bulk part of the hadron
wavefunction. Then the same is true for the bulk features (like
multiplicities and single--particle spectra) of particle production in
high--energy hadron--hadron collisions.

\texttt{(ii)} It exhibits {\em geometric scaling} : the occupation number
$n(k_{\perp},Y)$ depends upon the two kinematical variables $k_{\perp}$
and $Y$ only via the ratio $k_\perp^2/Q_s^2(Y)$. This property is very
interesting in that it is a consequence of saturation which manifests
itself {\em outside} the saturation region, at momenta\footnote{Namely,
it holds within a finite window above $Q_s$ whose with increases with $Y$
via BFKL diffusion \cite{Gelis:2010nm}.} $k_\perp \gg Q_s(Y)$. This
scaling transmits to the associated cross--sections and it has been
observed indeed, over a wide kinematical range, in the HERA data for DIS
at small $x\le 10^{-2}$ \cite{Stasto:2000er}. A similar scaling has been
predicted for MN jets in the asymmetric regime where the momenta
${k}_{1\perp}$ and ${k}_{2\perp}$ of the final jets are different (but
comparable) with each other and for sufficiently large rapidity
separations $Y\gtrsim 8$ \cite{Iancu:2008kb}.

The last point rises the question of how to compute hadronic
cross--sections in the presence of saturation. The $k_T$--factorization
has the same limitations as the BFKL equation: it holds only in the
dilute regime at sufficiently large momenta $k_\perp \gg Q_s(Y)$. Indeed,
if the gluon density becomes so high that the interactions among the
gluons in the wavefunction start to be important, then an external probe
which scatters off these gluons will undergo multiple interactions. So,
one needs a factorization scheme capable to include these interactions.
This is generally referred to as the {\em CGC (or high-energy)
factorization}, although its specific form (when known !) depends upon
the process at hand
\cite{Marquet,Albacete,Gelis:2010nm,Gelis:2008rw,Dominguez:2011wm}. For
DIS, one speaks about {\em dipole factorization} : the virtual photon
fluctuates into a quark--antiquark pair (a color dipole) which then
multiply scatters off the gluons in the target (see Fig.~\ref{Fig_CGC}
right). This multiple scattering is computed in the eikonal
approximation, by associating one `Wilson line' (a path--ordered
exponential of the target gluon field) to each of the two quarks
composing the dipole. Thus the DIS cross--section involves the product of
two Wilson lines averaged over the gluon fields in the target with the
CGC weight function. For the production of a pair of forward di-jets, the
cross--section is computed by taking the modulus squared of the amplitude
in Fig.~\ref{Pair} middle and involves four Wilson lines: two for the
partonic jets in the direct amplitude and two for those in the complex
conjugate amplitude.

One can now understand why the high--energy factorization cannot be
universal ({\em i.e.} process--independent), although there is some
systematics \cite{Dominguez:2011wm}: according to their partonic content,
different projectiles probe different correlation functions of the gluons
in the target. Moreover, these correlations naturally occur as products
of Wilson lines (one per parton) which encode multiple scattering. And
the evolution equations at high energy, as derived from the functional
JIMWLK equation, are most usefully written as coupled equations for
products of Wilson lines --- the Balitsky hierarchy. Besides, the whole
formalism (the evolution equations and the various factorization schemes)
is more conveniently written down in {\em impact parameter space} ---
that is, by using the transverse coordinates (${\bm x}_\perp$) instead of
the transverse momenta (${\bm k}_\perp$). Indeed, as shown by the example
of the Wilson lines, it is easier to deal with multiple interactions in
impact parameter space, where they exponentiate at the level of
$S$--matrix.

\begin{figure}[htb!]
  \begin{center}
    \includegraphics[width=0.36\textwidth]{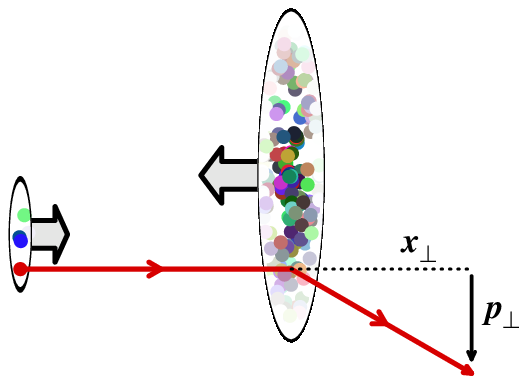}
    \hskip 5mm
    \includegraphics[width=0.54\textwidth]{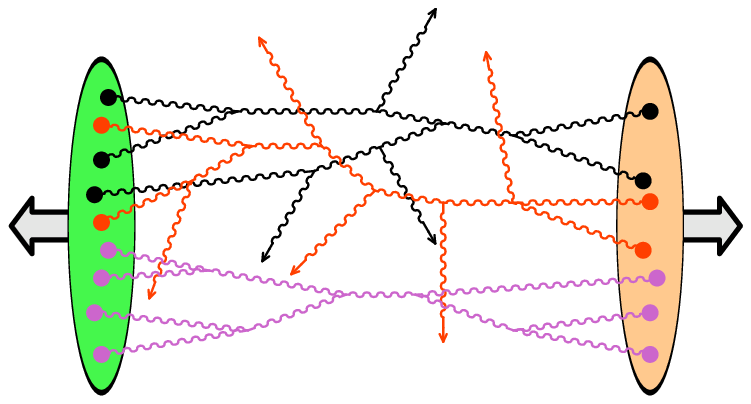}
  \end{center}
  \caption{  \label{fig:pA}\sl Left: sketch of a proton--nucleus
    collision. Right: multiparton interactions in a
    nucleus--nucleus collision.}
\end{figure}

In principle, the JIMWLK equation can be solved exactly, via numerical
methods; this amounts to computing a path integral which follows the
evolution with $Y$ in the functional space of the Wilson lines. The
feasibility of such a calculation has been demonstrated in
\cite{Rummukainen:2003ns}. But this procedure is rather tedious, so in
practice one generally relies on mean field approximations, notably the
BK equation, which is a non--linear equation for the dipole scattering
amplitude\footnote{The Fourier transform of the dipole amplitude can be
viewed as a natural extension of the unintegrated gluon distribution (to
which it reduces at $k_\perp \gg Q_s(Y)$) towards the saturation
regime.}, consistent with unitarity. There is by now a vast literature
devoted to applications of the BK equation to problems which admits a
dipole factorization, like DIS and the single--inclusive particle
production at forward rapidities. The state of the art, as reviewed by J.
Albacete \cite{Albacete}, involves solutions to the BK equation with
running coupling \cite{IancuRunBK}, which allows for successful fits to
the latest HERA data for the DIS structure function at $x\le 0.01$ and
any $Q^2$, and also to the RHIC data for forward particle production in
p+p and d+Au collisions.

\begin{figure}[htb!]
\begin{center}
\includegraphics[width=0.46\textwidth]{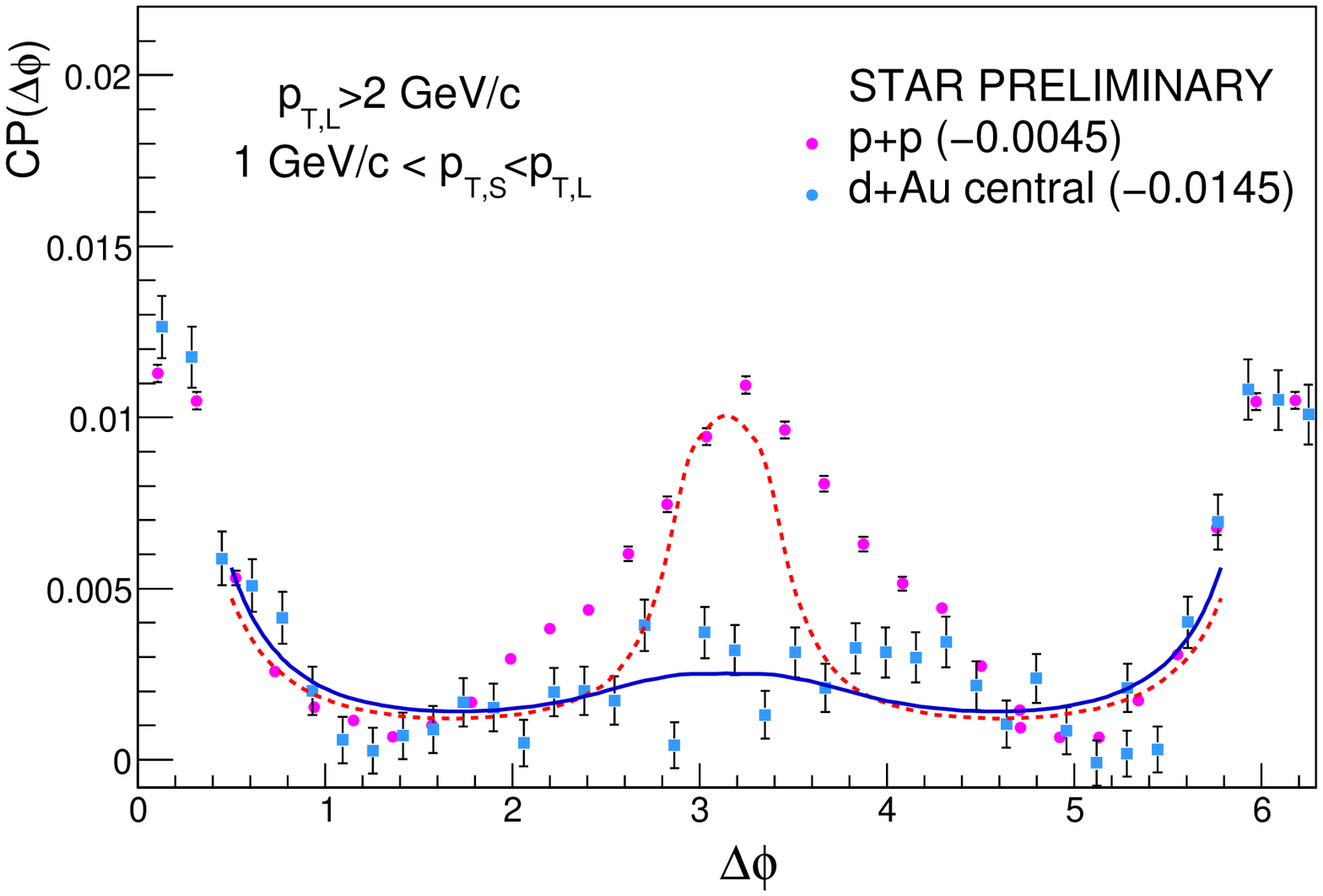}
\hskip 1mm
\includegraphics[width=0.49\textwidth]{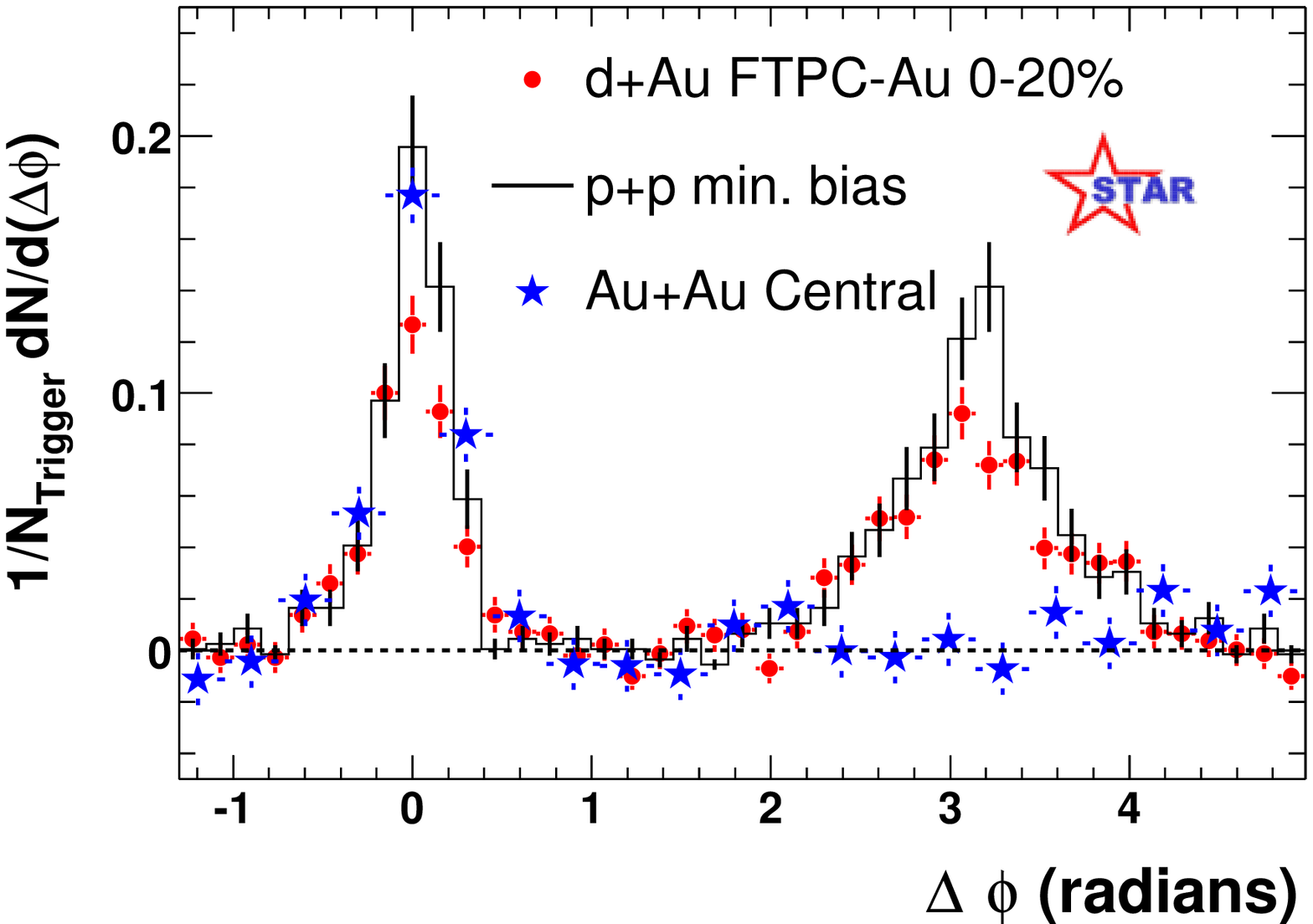}
\end{center}
\caption{\label{fig:DiJet}\sl Di--hadron correlations as measured by RHIC
(STAR) at forward (left) and central (right) rapidities.
Left: the leading (trigger) and the subleading particle
have rapidities $2.4< \eta_{1,2} < 4$. The `away' ($\Delta\phi=\pi$) peak
is visible for p+p collisions but is absent in d+Au. The
lines are theoretical calculations within the CGC framework
\cite{Albacete,Albacete:2010pg}. Right: all particles have $\eta\approx 0$
and $k_\perp= 2\div 6$~GeV. The `away' peak is clearly visible in p+p
and d+Au collisions, but is absent in Au+Au collisions. This is
`jet quenching' by interactions in the medium.}
\end{figure}

The d+Au collisions at RHIC are particularly favorable for studies of
saturation (see the left panel of Fig.~\ref{fig:pA}). The combination of
forward kinematics ($\eta=2.2\div 4$) with the large atomic number
$A=197$ enhances the saturation effects in the gold target. And the fact
that one can measure individual particles with relatively low
$k_\perp=1\div 3$~GeV up to rather large $\eta$ compensates for the
smaller COM energy (200~GeV per nucleon pair) as compared to the LHC.
Moreover, the target saturation momentum $Q_s(x)\simeq 1\,{\rm GeV}$ for
$x=10^{-4}$ is now comparable with the $k_\perp$ of the produced
particles, so the latter should be affected by saturation. An
experimental evidence in that sense is provided by the suppression of
particle production in d+Au collisions vs. p+p (the `nuclear modification
factor $R_{\rm d+Au}$' measured by BRAHMS \cite{Arsene:2004fa} and STAR
\cite{Adams:2006uz}), which becomes stronger and stronger with increasing
$\eta$ and also with increasing centrality. Both features are expected on
the basis of saturation, and the RHIC data are indeed nicely described by
fits using the BK equation with running coupling \cite{Albacete}. Another
piece of evidence in that sense has recently come from the measurement by
STAR \cite{Braidot:2010zh} of the azimuthal correlations in the
production of a pair of forward particles (cf. Fig.~\ref{Pair} middle)
with $\eta_{1,2} \simeq 3$ and $k_\perp\sim 2$~GeV: in d+Au collisions,
and unlike in p+p collisions, one sees `monojets' events where the `away'
peak at $\Delta\phi=\pi$ is totally absent (see the left panel in
Fig.~\ref{fig:DiJet}) ! Such a strong suppression is to be attributed to
saturation effects in the wavefunction of the target Au, as clear from
the fact that there is no similar phenomenon when the produced particles
have {\em central} ($\eta\approx 0$) rapidities (cf. the right panel in
Fig.~\ref{fig:DiJet}). And indeed the suppression seen at forward
rapidites can be qualitatively described by CGC--inspired calculations
\cite{Albacete:2010pg}, although a rigorous analysis (which would require
computing the correlation of a product of four Wilson lines) is still
lacking.

All the above examples where the high--energy factorization is well under
control refer to a {\em dense--dilute scattering} ---  the situation
where the projectile is relatively dilute and can be described as a
collection of a few partons, whereas the target is dense (possibly at
saturation) and is described as a CGC ---, and to a relatively {\em
simple final state} --- the inclusive production of a few partons. Some
more complicated problems are the {\em dense--dense scattering}, e.g. the
collision between two heavy ions to be discussed in the next section, and
the description of exclusive final states in the presence of saturation
and multiple scattering. In what follows, I would like to describe two
methods for the construction of the final state in a high--energy
scattering that have been presented at ISMD2010.

The first method, as described by H. Jung \cite{Jung} and M. Deak
\cite{Deak}, uses $k_T$--factorization together with a linear, BFKL--like
evolution of the unintegrated gluon distributions to describe the hard
scattering (say the production of a pair of jets) and identifies the
`underlying event' (the relatively soft radiation accompanying the hard
jets) with the gluons from the partonic cascades involved in the
scattering; these gluons are assumed to be liberated in the collision.
For this description to be more realistic, it replaces the BFKL evolution
with the CCFM one
--- a generalization of BFKL which takes into account the {\em angular
ordering} along the partonic cascade. The angular ordering follows from
quantum interference between successive emissions, which implies that any
newly emitted gluon should make a larger angle with the collision axis
than all the gluons emitted before it. (A similar property holds for the
radiation produced via jet fragmentation in the final state; but in that
case, the angles are decreasing from one emission to the next one, rather
than increasing.) This whole scheme has been numerically implemented in
the Monte--Carlo generator CASCADE, with results which satisfactorily
describe the small--$x$ phenomenology at HERA and which predict harder
$k_\perp$--spectra for the forward jets at the LHC as compared to more
standard event generators like PYTHIA. Note that there is no saturation,
nor multiple interactions, in this approach, although one could mimic
saturation by introducing an energy--dependent infrared cutoff
(`saturation boundary'), which ideally should be self--consistently
computed within the CCFM evolution \cite{Avsar:2009pf}.

\begin{figure} [htb!]
\centerline{
 \includegraphics[width=0.55\textwidth]{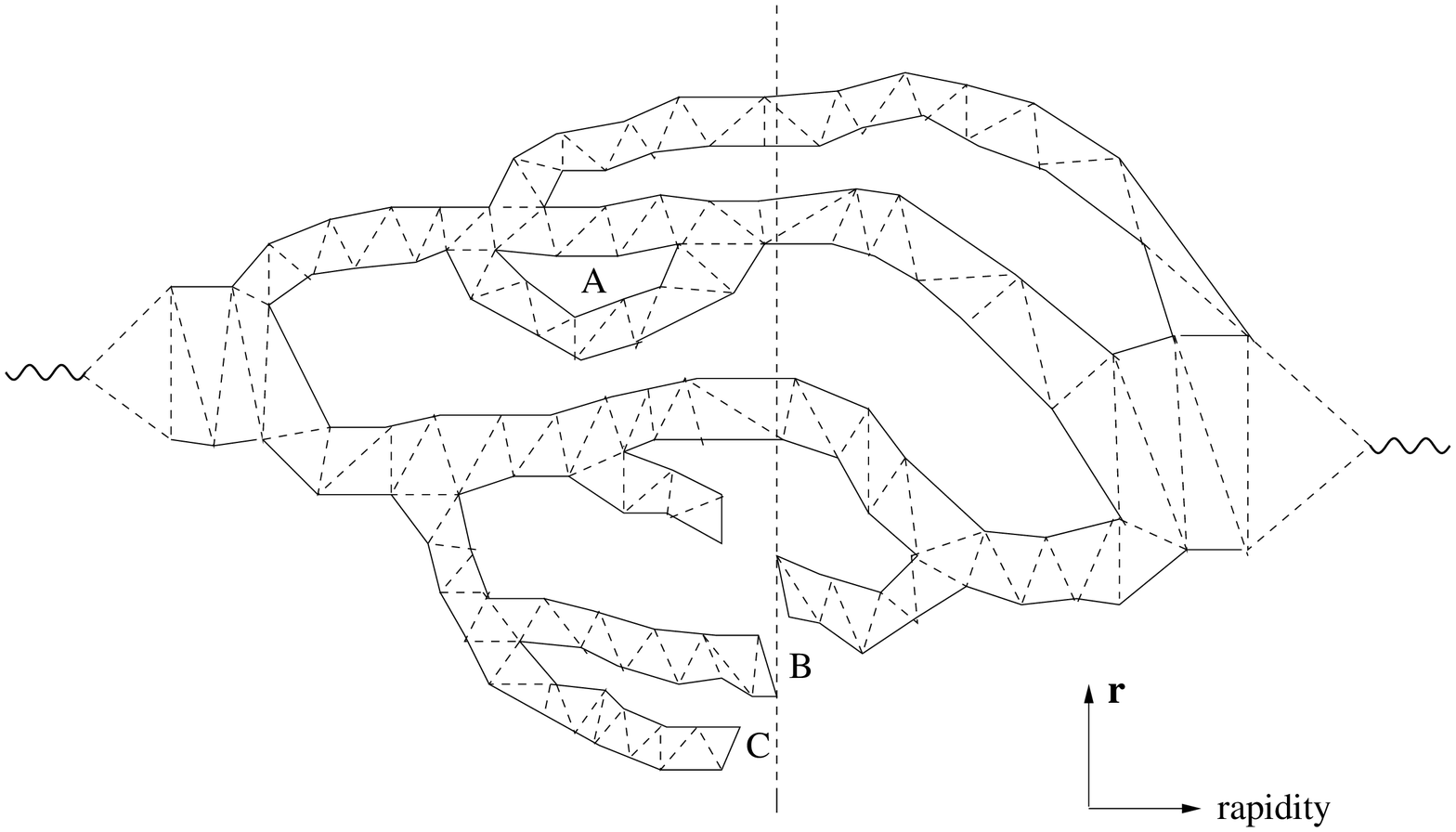}
 \includegraphics[width=0.45\textwidth]{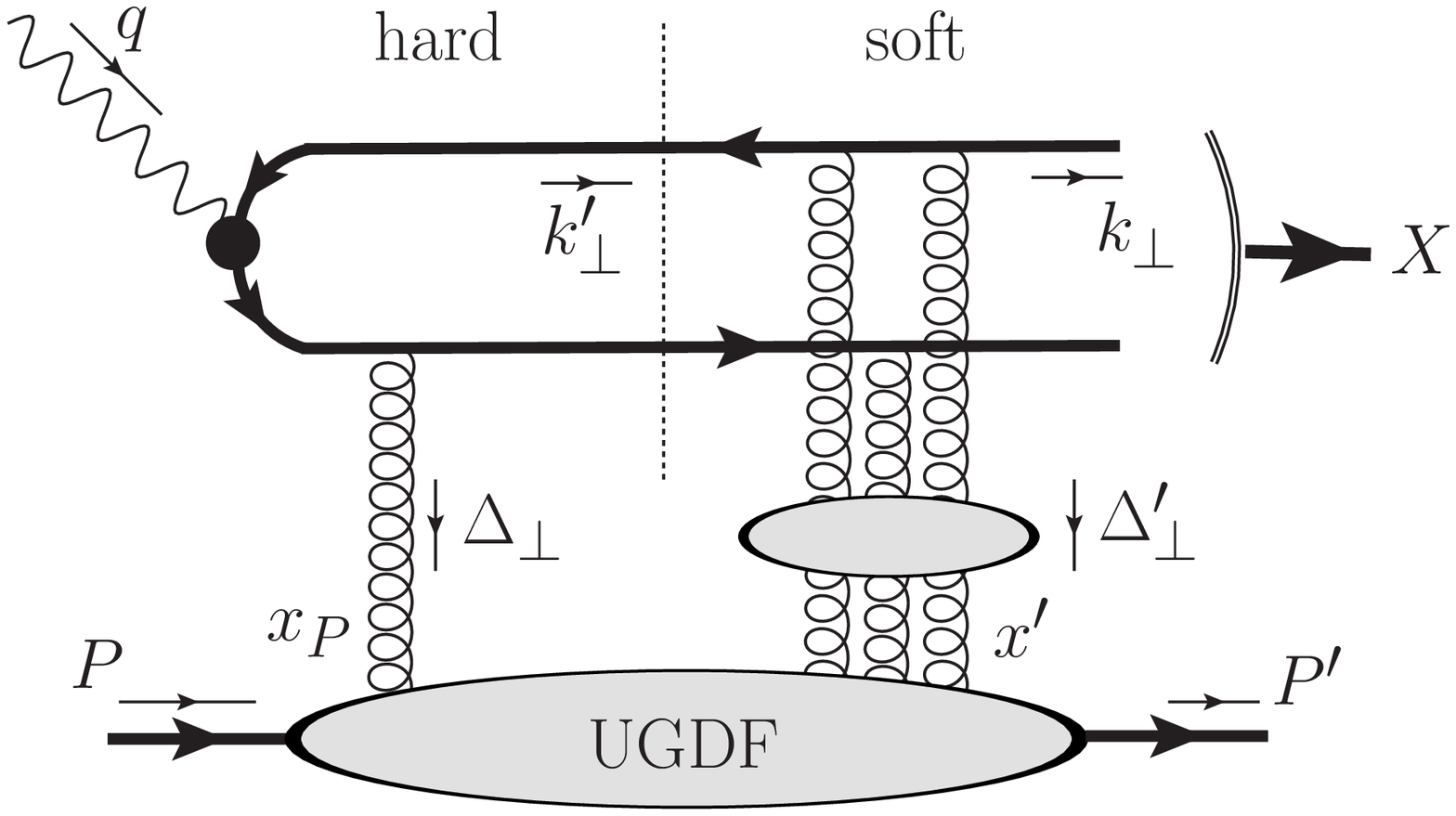}}
 %\vspace*{-.6cm}
\caption{\sl Left: A
collision between two evolved dipole cascades in the DCM~\cite{Gustafson}.
One sees 3 subcollisions and a Pomeron loop ($A$)
which opens via dipole splitting and closes via a dipole swing.
Right: The `hard+soft' model for diffractive DIS~\cite{Enberg}.
\label{Fig_DCM}}
\end{figure}

A very ambitious program, which aims at explicitly including saturation
and multiple scattering together with the correlations induced by the
high--energy evolution, has been presented by G. Gustafson
\cite{Gustafson}. This approach is based on an extension of Mueller's
dipole picture, which is a reformulation of the LO BFKL evolution, valid
at large $N_c$, where gluons are replaced by color dipoles which multiply
via $1\to 2$ dipole splitting. By itself, this picture cannot accommodate
saturation (there is no $2\to 1$ dipole recombination in pQCD !), but it
has the virtue to capture the fluctuations and correlations produced by
the BFKL evolution in the dilute regime. Such fluctuations determine the
cross--section for diffractive excitations in the Good--Walker picture.
They may also be important for the evolution towards saturation, as
suggested by the correspondence between the high--energy evolution in QCD
and the reaction--diffusion process in statistical physics\footnote{Note
that the BK equation \eqref{BK} is essentially the same as the FKPP
equation describing `reaction--diffusion' in the mean field approximation
\cite{iancuFKPP} (and refs. therein).} \cite{iancuFKPP}. In fact, in a LO
BFKL picture, these fluctuations are truly huge and rapidly growing with
$Y$ \cite{iancuSalam95,Avsar:2008ph}. In real QCD, though, they are
considerably tamed by saturation in the high--density regime and by the
running of the coupling in the dilute regime \cite{Dumitru:2007ew}. So
far, there is no first--principle theory for the QCD evolution including
both fluctuations and saturation (and hence {\em Pomeron loops}), but
some semi--heuristic approaches have been developed in that sense
\cite{iancuPloop}, including the Dipole Cascade Model by Gustafson and
collaborators \cite{iancuSwing}. In this approach, saturation is
introduced via `dipole swing' --- a rule for color reconnections which
prohibits the formation of large dipoles with transverse size $r\gtrsim
1/Q_s(Y)$. Also multiple sub--collisions are permitted in the interaction
between two evolved dipole cascades, as illustrated in Fig.~\ref{Fig_DCM}
(left). The results of the Monte--Carlo implementation of this model
nicely illustrate the role of saturation and multiple interactions in
suppressing fluctuations and removing unitarity violations, which would
otherwise be substantial in p+p collisions at the LHC energies. The
extension of the model to exclusive final states has been very recently
given~\cite{Flensburg:2011kk}.

R.~Enberg~\cite{Enberg} has presented a new, `hard+soft', model for
diffractive deep inelastic scattering, which compares quite well with the
respective HERA data for $Q^2\ge 10$~GeV$^2$. The diffractive process is
interpreted as the consequence of a hard scattering followed by a series
of soft gluon exchanges, resummed in the eikonal approximation, resulting
in an overall color singlet exchange (see Fig.~\ref{Fig_DCM} right). This
resummation is similar to that performed by Wilson lines in the CGC
formalism, so it would be interesting to clarify the interplay between
these two approaches.

\section{High density: the quest for the QGP}
\label{CGC}

The CGC effective theory also provides the {\em initial conditions} for
ultrarelativistic heavy ion collisions (HIC), as measured at RHIC and the
LHC. By `initial conditions' I mean the wavefunctions of the incoming
nuclei (which at high energy are dominated by small--$x$ gluons) and the
particle (actually, parton) production during the early stages of the
collision, up to times $\tau_0\simeq 1/Q_s\simeq 0.1\div 0.2$~fm. Here
$Q_s$ is the average saturation momentum in any of the two colliding
nuclei, as probed by multiparticle production at central rapidities. Most
of the particles produced in a A+A collision are `minijets' with
semi-hard transverse momenta $k_\perp\sim 1$~GeV, of the same order as
$Q_s$ at the relevant values of $x$ (from $10^{-3}$ to $10^{-4}$). This
is not a coincidence: these particles are either partons (mostly gluons)
from the initial wavefunctions that have been liberated by the collision
(via multiple short--range scattering among partons), or the products of
their subsequent fragmentation. But after being liberated, the density of
these partons within the interaction region is so high, and the size and
lifetime of this dense partonic system are so large, that multiple
scattering will play an important role in redistributing the energy and
momentum and driving the system towards (local) thermal equilibrium. One
can appreciate the density of this system either by estimating its energy
density --- one finds $\varepsilon \gtrsim 15$~GeV/fm$^3$ at the LHC,
which is about 10 times larger than the density of nuclear matter (and 3
times larger than in Au+Au collisions at RHIC) ---, or from the measured
multiplicity in the final state --- about 1600 particles per unit
rapidity in central Pb+Pb collisions with $\sqrt{s_{_{NN}}} = 2.76$~TeV
at the LHC \cite{Aamodt:2010pb}.

\begin{figure}[htb!]
\centerline{
 \includegraphics[width=0.8\textwidth]{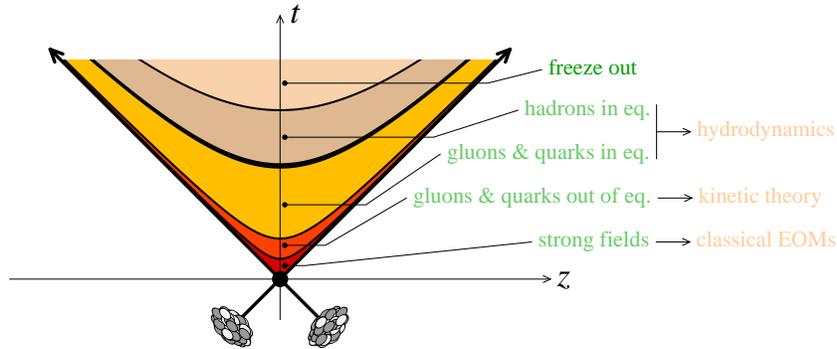}}
 %\vspace*{-.6cm}
\caption{\sl The expected space--time picture for a
ultrarelativistic heavy ion collision.
\label{Fig_HIC}}
\end{figure}

The `standard scenario' for the evolution of this dense partonic system
until hadronization, as emerging from theoretical considerations amended
by the experimental reality at RHIC, is illustrated in
Fig.~\ref{Fig_HIC}. There are still many zones of shadow in this scenario
--- notably, about the precise time scales, the mechanism responsible for
thermalization, and the nature of the {\em quark--gluon plasma} (QGP)
which gets formed in this way (a `weakly interacting gas' or a `nearly
perfect fluid' ?) --- and many interesting observables whose theoretical
interpretation is still under debate --- like the `ridge effect' or the
strong `jet quenching'. In what follow I shall summarize the way how such
open questions have been reflected in the talks and discussions at
ISMD2010, by following the chronology of the HIC as illustrated in
Fig.~\ref{Fig_HIC}.

The calculation of particle production in the early stages of a heavy ion
collision is particularly complex because of the need to account for two
types of multiparton interactions: gluon saturation in the wavefunctions
of the incoming nuclei and multiple subcollisions between partons in the
two nuclei (see the right panel of Fig.~\ref{fig:pA}). Within the CGC
formalism, two strategies have been devised to perform such calculations,
with different levels of rigor \cite{Gelis:2010nm}:

(a) A heuristic extension of the $k_\perp$--factorization, which includes
saturation effects within the `unintegrated gluon distributions' of the
two nuclei (obtained as solutions to the BK equation), but neglects the
multiple scattering in A+A collisions. In principle, such a strategy is
justified only for sufficiently large momenta $k_\perp\gg Q_s$. In
practice, this has been quite successful in describing bulk features of
the particle production like the multiplicity density $\rmd N/\rmd \eta$
at $\eta=0$, which involve an integration over all values of $k_\perp$
\cite{Albacete}. Such calculations predict $\rmd N/\rmd\eta \sim Q_s^2
R_A^2$, and hence a power--law increase of the multiplicity with the
center of mass energy: $\rmd N/\rmd \eta\sim {s}^{\lambda/2}$ with
$\lambda\approx 0.25$. This is in agreement with the observed rise in the
multiplicity when going from RHIC to the LHC \cite{Aamodt:2010pb}.
Another interesting consequence of such calculations, as explained by
G.~Wolschin~\cite{Wolschin} at ISMD2010, is an increased baryon stopping
due to saturation: the (large--$x$) valence quarks from one nucleus
scatter off the high-density, small--$x$, gluons from the other one and
thus get redistributed in rapidity; in particular, some of them are
slowed down to smaller rapidities, which increases the net baryon density
around $\eta=0$ and, especially, pushes the fragmentation peaks towards
smaller values of $\eta$. This effect should be strong enough to allow
studies of saturation at the LHC based on the position of the
fragmentation peaks.

(b) The CGC factorization \cite{Gelis:2008rw,Gelis:2010nm}, in which the
two nuclei prior to the collision are described as statistical ensembles
of strong, classical, color fields, and the collision is represented by
the solution to the classical Yang--Mills equations with initial
conditions randomly chosen from these ensembles. This includes both
saturation (via the CGC weight functions for the two nuclei) and multiple
scattering (via the non--linear effects in the Yang--Mills equations) and
is well adapted for the computation of sufficiently inclusive quantities
like the single--particle spectra or the energy density and its
correlations. But it has the drawback to require heavy numerical
simulations, in the form of classical lattice calculations.

\begin{figure}[htb!]
\begin{center}
\includegraphics[width=0.4\textwidth]{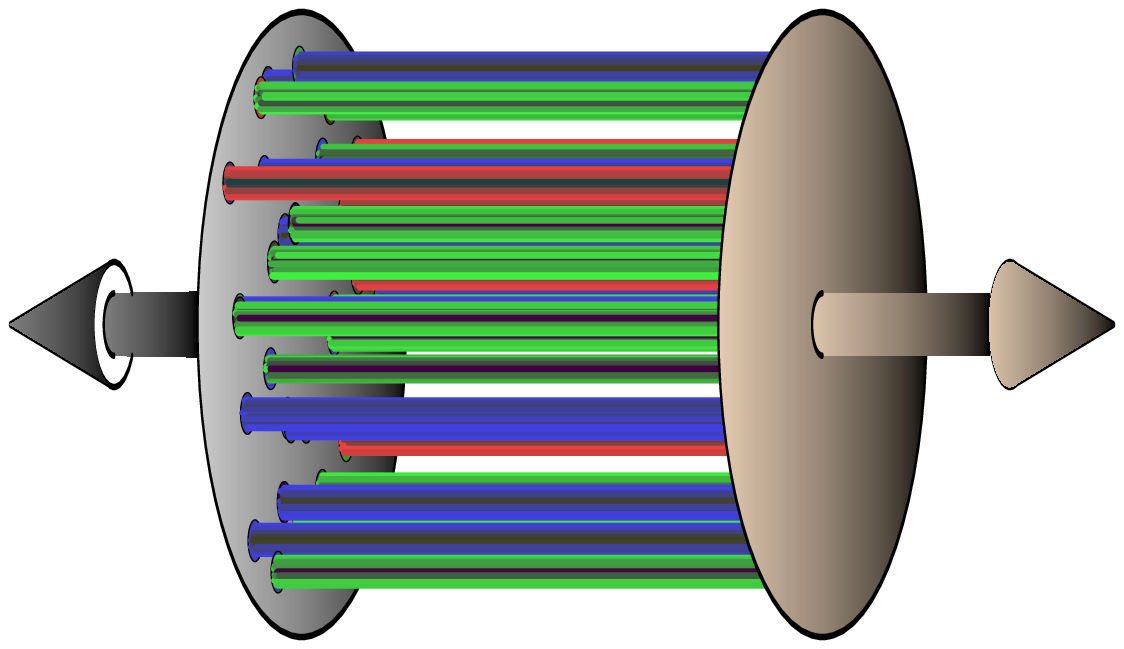}
\hskip 1mm
\includegraphics[width=0.5\textwidth]{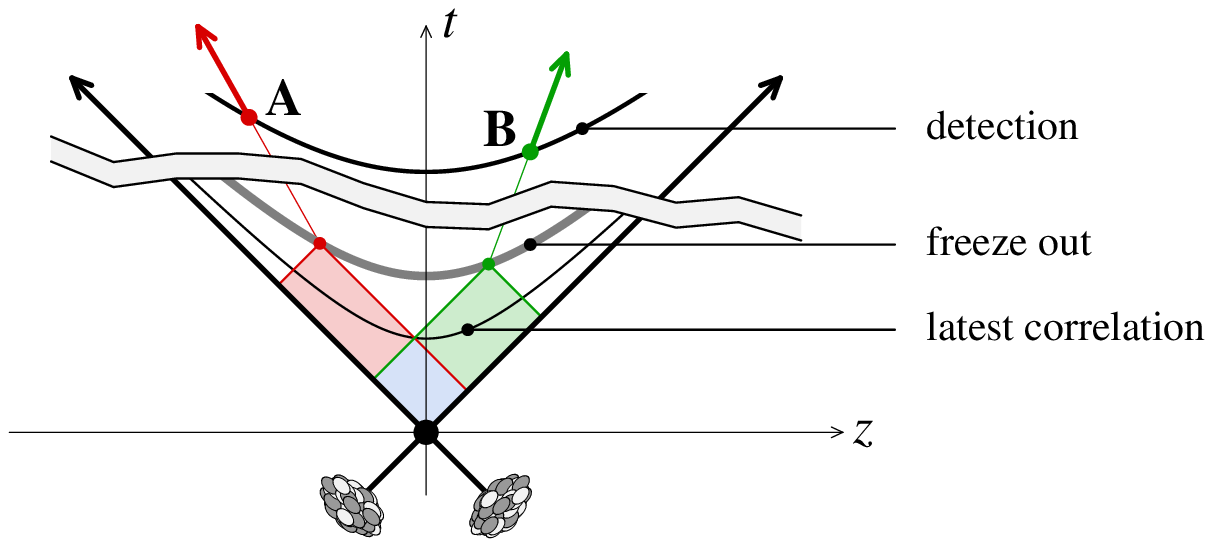}
\end{center}
\caption{\label{fig:flux-tubes}\sl Left: Gauge
  field configurations in the form of ``flux tubes'' of longitudinal
  chromo-electric and chromo-magnetic fields screened on transverse
  scales $1/Q_s$. Right: causal relations
  between two particles separated in rapidity.}
\end{figure}

The non--equilibrium, dense, partonic matter produced in the early stages
($\tau_0 \sim 1/Q_s$) of a A+A collision is called the `Glasma'. The
solutions to the Yang--Mills equation alluded to above show that the
Glasma fields are longitudinal chromo-electric and chromo-magnetic fields
with occupation numbers $\sim 1/\alpha_s$, that are screened at distances
$~1/Q_s$ in the transverse plane of the collision. As a consequence, the
matter produced can be visualized (see Fig.~\ref{fig:flux-tubes}) as
comprising $R_A^2 Q_s^2$ color flux tubes of size $1/Q_s$, each producing
$1/\alpha_s$ particles per unit rapidity.

The Glasma flux tubes carry topological charge~; the resulting,
dynamical, topological transitions (`sphalerons') may result in
observable metastable CP--violating domains. As explained by
H.~Warringa~\cite{Warringa} at ISMD2010, such transitions could explain
some puzzling charge correlations observed by STAR, via the `chiral
magnetic effect'~: under the action of the ultra strong magnetic fields
($B\sim 10^{18}$~Gauss) which are likely to be created in peripheral
ultrarelativistic A+A collisions, the fluctuations in the topological
charge can induce fluctuations in the electric charge density in the
final state.

\begin{figure}[htb!]
\begin{center}
\includegraphics[width=0.45\textwidth]{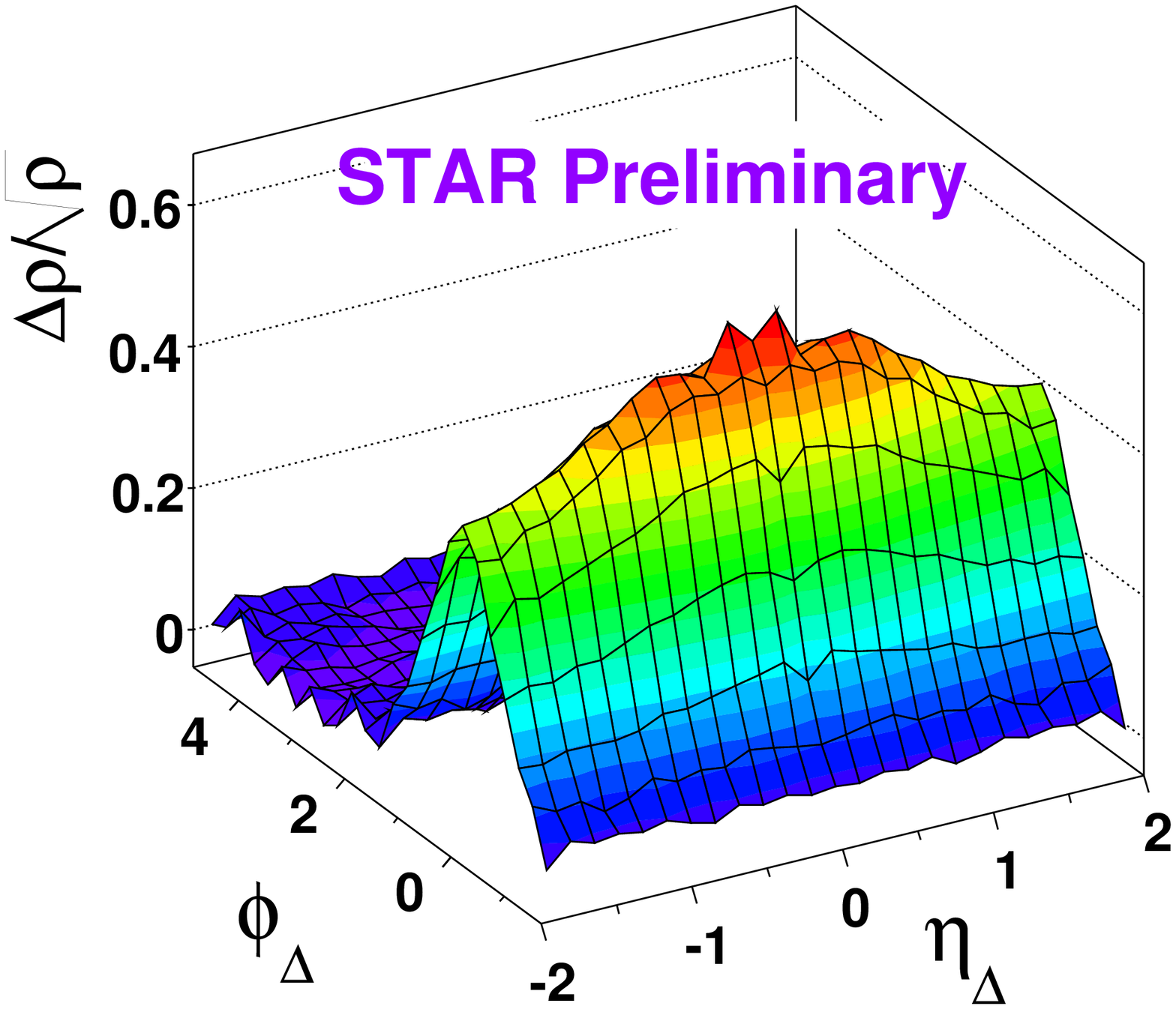}
\hskip 1mm
\includegraphics[width=0.5\textwidth]{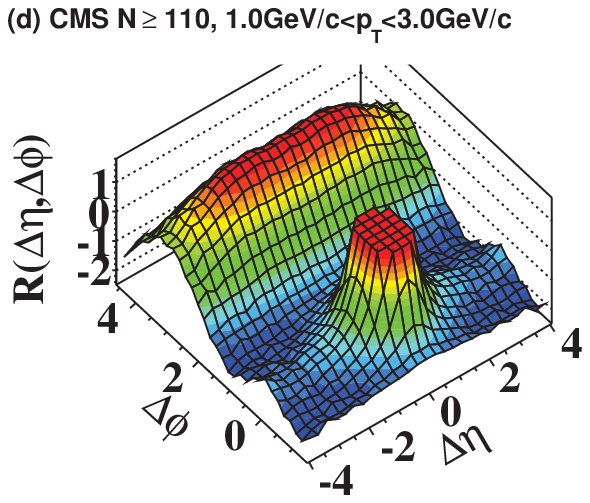}
\end{center}
\caption{\label{fig:ridge}\sl Ridge effect in Au+Au collisions
RHIC~\cite{iancuridgeRHIC} (left) and in high--multiplicity events in
p+p collisions at the LHC~\cite{Khachatryan:2010gv} (right).}
\end{figure}

The Glasma flux tubes also generate $n$--particle long--range rapidity
correlations and thus can naturally explain the {\em ridge effect} seen
in di--hadron correlations in Au+Au collisions at
RHIC~\cite{iancuridgeRHIC} and also (with a much lower intensity, though)
in p+p collisions at the LHC~\cite{Khachatryan:2010gv}. The `ridge' is a
two--particle correlation in the distribution of particles accompanying a
jet which extends with nearly constant amplitude over several units in
rapidity ($\Delta\eta\simeq 4$) and which is well collimated in the
azimuthal separation $\Delta\varphi$ relative to the jet --- so it looks
like an extended mountain ridge in the $\Delta \eta$--$\Delta \varphi$
plane (see Fig.~\ref{fig:ridge}). This means that particles which
propagate along very different directions with respect to the collision
axis preserve nevertheless a common direction of motion (close to that of
the triggered jet) in the transverse plane. By causality, such a
correlation must have been induced at early times, when these particles
--- which rapidly separate from each other --- were still causally
connected (see the right panel of Fig.~\ref{fig:flux-tubes}).

The ridge in A+A collisions can be explained by the transverse radial
flow of the particles produced by the decay of the Glasma flux tubes
\cite{Dumitru:2008wn}. These particles are correlated in rapidity over a
range $\Delta\eta\sim 1/\alpha_s$ since the flux tubes are uniform in
$\eta$ over that range. The angular collimation occurs because particles
produced isotropically in a given flux tube are collimated by their
subsequent radial, outward, flow \cite{Voloshin:2003ud}. This collimation
effect is clearly demonstrated by the hydrodynamical calculations
presented at ISMD2010 by R. Andrade~\cite{Andrade} and Y.
Hama~\cite{Hama}. In the case of p+p collision, where no flow is
expected, a small collimation may be generated by the di--hadron
production mechanism in the presence of saturation \cite{Dumitru:2010iy}.
This would be in agreement with the experimental observation by the
CMS~\cite{Khachatryan:2010gv} that the ridge is visible only in the
high--multiplicity (central ?) events and only within a limited range in
$k_\perp$ (from 1 to 3~GeV), which is in the ballpark of the proton
saturation momentum at the LHC.

The evolution of the Glasma into a thermalized Quark Gluon Plasma (QGP)
is not yet fully understood, but some of the RHIC  data --- notably the
collective motion known as {\em elliptic flow} --- suggest that this
happens very fast, on a time scale of one fermi/c. Hydrodynamical
calculations were quite successful in explaining the elliptic flow seen
at RHIC \cite{Back:2004je,Adcox:2004mh} and predicting the one recently
observed by the LHC \cite{Aamodt:2010pa}, but to that aim they had to
assume a very early thermalization time $\tau_{\rm eq} \lesssim 1$~fm/c
and a relatively low viscosity. Explaining such a rapid thermalization
from first principles is a challenge for the theory, and so is also the
identification of new observables which would permit a more refined study
of the approach towards thermalization in the data. In particular, we
would like to know how fast is this plasma evolving towards {\em
isotropy} from the Glasma initial conditions, which are highly
anisotropic. To address such questions, W.~Florkowski and
collaborators~\cite{Florkowski} have developed a generalized
hydrodynamical framework (ADHYDRO) which can accommodate highly
anisotropic initial conditions and strong dissipation (proportional to
the anisotropy). So far, calculations have been performed for the
one--dimensional, longitudinal, expansion, with results which show the
expected approach towards isotropy, on a time scale $\tau_{\rm iso}$
which is a parameter of the model. Within a related approach,
P.~Bozek~\cite{Bozek} concluded that this isotropisation time should be
very small, $\tau_{\rm iso}\lesssim 0.25$~fm/c, in order to reproduce the
relatively large directed flow $v_1$ seen in the RHIC data: increasing
$\tau_{\rm iso}$ would rapidly kill $v_1$. Hence, the directed flow is a
very sensitive probe of thermalization, unlike the observables related to
the transverse dynamics, like the $k_\perp$ spectra or the elliptic flow
$v_2$, which are only little affected by the initial pressure anisotropy.

To be able to identify and study the QGP in HIC, one needs a good
comprehension of its properties. For the case of a plasma in thermal
equilibrium and with zero fermionic density, lattice QCD provides such a
comprehension from first principles. In his review talk of this topic at
ISMD2010, Z.~Fodor~\cite{Fodor} emphasized that, due to tremendous
progress in numerical and computational techniques and to the strenuous
efforts of continuously growing and re-concentrating collaborations,
lattice QCD has finally reached the `productive phase', where the
continuum limit and the finite--size scaling are under control, and so is
also the extrapolation to physical masses for the lightest quarks. This
makes it possible to have a good control of the hadron spectrum with only
few input parameters and, in particular, `predict' the proton mass with
high accuracy. The same progress allowed one to clarify the nature of the
deconfinement `phase transition' --- which is actually a smooth
cross--over, as demonstrated by the smooth behaviour of the
Polyakov--line susceptibility in the thermodynamic limit $V\to\infty$
(approached on the lattice via finite--size scaling) --- and to solve a
longstanding controversy concerning the value of the  `critical'
temperature for deconfinement, for which a value $T_c\approx 175$~MeV
seems to be now widely accepted.

The lattice studies of the equation of state~\cite{Fodor} also reveal
that, after a rapid increase around $T_c$, the energy $\varepsilon$ (or
entropy $s$) density of the QGP is very slowly approaching the
Stefan--Boltzmann limit, in such a way that $s_{\rm QCD}(T)\simeq
(0.80\div 0.85)\,s_{\rm SB}(T)$ when $T$ varies from 2 to 5 $T_c$ (the
temperature range of interest for HIC at RHIC and the LHC). Such a
deviation of less than 20\% from the ideal gas limit is small enough to
suggest a weak--coupling behaviour, and it is indeed well accounted by
calculations using `Hard Thermal Loop' resummations of the perturbative
expansion at finite temperature \cite{Blaizot:2003tw}. These calculations
support the picture of the QGP as a weakly--interacting gas of
`quasiparticles', quarks and gluons, which are dressed by medium effects
in a way that is computable in perturbation theory.

On the other hand, this value $s_{\rm QCD}/s_{\rm SB}\approx 0.80$ is
also close to the value $s_{\infty}/s_{\rm SB}=0.75$ predicted by the
AdS/CFT correspondence \cite{iancuAdSCFT} for the strong coupling limit
$g^2N_c\to\infty$ of ${\mathcal N}=4$ supersymmetric Yang--Mills theory
--- a `cousin' of QCD which has the color gauge symmetry SU$(N_c)$, but
also (four) sypersymmetries, and which is conformal at quantum level (the
coupling is fixed). Hence, by themselves, the lattice QCD results for the
thermodynamics cannot exclude the possibility that the QGP be {\em
strongly coupled} in this range of temperatures\footnote{The coupling
constant $\alpha_s=g^2/4\pi$ in QCD can never become arbitrarily large,
because of asymptotic freedom, but it can be of order one at scales of
order $\Lambda_{\rm QCD}$ and this might lead to an effectively
strong--coupling behaviour. In fact, for $2T_c< T < 5T_c$, $g^2N_c\simeq
7\div 10$ is indeed quite large.}. Why would be such a possibility
interesting~? As earlier mentioned, a successful hydrodynamical
description of the RHIC data for the elliptic flow requires a short
thermalization time and a very low viscosity--to-entropy ratio
$\eta/s=0.1\div 0.2$, two properties which are rather difficult to
explain at weak coupling, but which become natural if the coupling is
strong. (Recall, e.g., that in kinetic theory, the viscosity is
proportional to the mean free path, which decreases with increasing
coupling.) And indeed AdS/CFT calculations in gauge theories with a
gravity dual, like ${\mathcal N}=4$ SYM, predict a small, universal,
lower bound $\eta/s=1/4\pi$ in the strong coupling limit
\cite{iancuSonStar}, which is compatible with the phenomenology at RHIC
\cite{Luzum:2008cw}. Although such calculations refer to conformal field
theories, it still make sense to compare their results to the QGP phase
of QCD, since from lattice QCD we know that the `trace anomaly' (the
violation of conformality by the running of the coupling) is quite small
for temperatures $T\gtrsim 2T_c$ \cite{Fodor}.

A strong--coupling scenario could also explain the relatively strong {\em
jet quenching} observed in A+A collisions at RHIC
\cite{Adams:2005dq,Adcox:2004mh} and the LHC
\cite{Aad:2010bu,Aamodt:2010jd,Collaboration:2011sx}. This refers to the
suppression of particle production with respect to `naive' extrapolations
from p+p collisions, as measured by the nuclear modification factor
$R_{\rm A+A}$, or by the suppression of the `away' jet in di--hadron
correlations (see the right panel in Fig.~\ref{fig:DiJet}). Such
phenomena indicate that the deconfined medium created in the intermediate
stages of a HIC is very {\em opaque}, including for relatively hard
probes with $k_\perp=2\div 20$~GeV. Within pQCD, the dominant mechanism
for parton energy loss in the medium is gluon radiation stimulated by the
scattering off the medium constituents \cite{Wiedemann:2009sh}. But this
mechanism seems unable to explain the strong suppression seen at RHIC,
although definitive conclusions cannot be drawn, given the difficulty to
perform realistic calculations. At strong coupling, AdS/CFT
calculations~\cite{Iancu:2008sp,Gubser:2009sn,CasalderreySolana:2011us}
suggest that a new mechanism for energy loss should come into play
--- the medium--induced parton branching \cite{Iancu:2008sp}. It is
likely that all these mechanisms will coexist for realistic values of the
coupling. Interestingly, the RHIC data for elliptic flow (which refer to
relatively soft particles with $k_\perp\sim 1$~GeV) and those for jet
quenching (where $k_\perp\simeq 5$~GeV is semi--hard) can be
simultaneously accommodated by models inspired by the respective AdS/CFT
predictions at strong coupling, as explained by J.~Noronha~\cite{Noronha}
at ISMD2010.

With the advent of the LHC, it became possible to perform calorimetric
measurements of the energy loss for {\em real} jets (as opposed to
leading particles). The first results in that sense
\cite{Aad:2010bu,Collaboration:2011sx} are already impressive: for
central Pb+Pb collisions and for very hard jets with $E_T\ge 100$~GeV,
one sees highly asymmetric dijet events, characterized by a large energy
imbalance (a few dozens of GeV) between two back--to--back jets. This
rises the challenge of understanding jet evolution (fragmentation and
energy loss) in a dense medium --- a topic addressed by T.~Trainor at
ISMD2010~\cite{Trainor}. It remains as an interesting open question
whether these new results at the LHC can be fully explained by weak
coupling calculations (as one may expect for such hard jets), or if there
is still place for non--perturbative phenomena associated with the
interactions between the radiation and the medium.

The QGP created in a heavy ion collision keeps expanding and hence it
cools down, until the (local) density becomes so low that it must
hadronize. The abundances of strange and non-strange mesons and baryons
produced in heavy ion collisions in a wide range of collision energies
are consistently described by statistical physics within the `hadron
resonances gas' model --- that is, as an ideal gas of hadrons with
`freeze-out' temperatures and baryon chemical potentials that are a
function of collision energy only. Moreover, the freeze-out temperature
extracted at the highest RHIC energy, of about 170~MeV (at zero baryon
chemical potential), is very close to the critical temperature for
deconfinement from lattice QCD. In his contribution to ISMD2010,
J.~Cleymans~\cite{Cleymans} has reviewed the status of chemical
equilibration in HIC at the freeze-out. In particular, he has emphasized
that the net baryon density at the freeze-out, which is small for both
very high (RHIC) and very low (FAIR) energies, should exhibit a maximum
value $\rho_0\simeq 0.15$~fm$^{-3}$ corresponding to a freeze-out
temperature $T\sim 140$~MeV.

\section{Soft interactions: the quest for a theory}
\label{Soft}

The physics of soft interactions at high energy --- as relevant, e.g.,
for the calculation of the total, elastic, and diffractive
cross--sections in hadron--hadron collisions, or for the description of
the underlying event --- has been the main scope of the early ISMD
meetings and a main focus for all its subsequent editions over the last
forty years. But in spite of undeniable progress, this remains the topics
for which we have the less satisfactory understanding from first
principles. There is, of course, a `good' reason for that: this topics
lies on the `dark' (non--perturbative) side of QCD, and unlike other
related topics like hadron spectroscopy it cannot be addressed via
lattice calculations. The progress in this field can be associated with
two main directions: \texttt{(i)} a shift in the borderline between {\em
Terra Incognita} (the non--perturbative sector of QCD) and {\em Mare
Nostrum} (the pQCD `sea' that we know how to navigate over), and
\texttt{(ii)} the development of new models, more sophisticated and
better motivated, which heuristically describe some of the `dark regions'
of {\em Terra Incognita}.

With increasing energy, many of the phenomena that were traditionally
associated with soft interactions are in fact controlled by the
semi--hard ones. As discussed in the previous sections, the wavefunction
of an energetic hadron is dominated by small--$x$ gluons with transverse
momenta of the order of the saturation momentum, which is semi--hard
($Q_s\sim 1$~GeV) at the current energies. These gluons are liberated in
a high--energy collision and they form the bulk of the `underlying event'
(UE) --- the ensemble of radiation accompanying a hard parton--parton
interaction and which are not directly associated with that interaction.
After being liberated, the gluons can still evolve in the `final state',
via multiparticle interactions (which become important at high energies),
fragmentation (radiation of new quanta), and hadronization. Some of these
processes can still be treated in perturbation theory. And the genuinely
soft ones, like hadronization, should not affect observables like the
energy flow or the single--particle spectra (although they are of course
essential for rapidity gaps and diffraction). Thus, besides the hard
scattering, there is a significant part of the UE which in principle can
be described within pQCD. For the remaining, non--perturbative, part one
has to resort on models involving free parameters, like the Lund string
model for hadronization.

Such a hybride description of the final state, combining perturbative and
non--perturbative ingredients, has been implemented in Monte--Carlo (MC)
event generators~\cite{Buckley:2011ms}, which in practice are quite
successful even though they do not fully reflect our current fundamental
understanding (on the pQCD side). For instance, the non--linear physics
associated with gluon saturation in the initial wavefunctions and with
multiparticle interactions in the final state is not properly included,
but only mimicked by the introduction of an energy--dependent cutoff
$p_0(\sqrt{s})$ (actually, an Ersatz of $Q_s(x)$) separating `hard' and
`soft' interactions, and by the heuristic resummation of multiple
scattering in the eikonal approximation. This treatment neglects
coherence phenomena expected at high density and thus overestimates the
cross--sections for the production of `minijets'. To compensate for that,
the separation scale $p_0(\sqrt{s})$ must be allowed to grow very fast
with the energy and thus become even larger than the actual saturation
scale ($p_0\simeq 5$~GeV at the LHC~\cite{Deng:2010mv}). In principle,
one could avoid such heuristic short-cuts and rely on the CGC formalism
for a proper description of the transition region around $Q_s$. In
practice, however, this formalism seems difficult to reconcile with the
conventional MC generators.

To understand this difficulty, one should remember that, in general, the
predictions of quantum mechanics cannot be reproduced by a classical
stochastic process, so like a Monte-Carlo. This only works under specific
approximations, which neglect interference effects (or treat them only
approximately) and whose applicability domains are limited and often
mutually exclusive. Two examples in that sense are the collinear
factorization at high $Q^2$, which lies at the basis of most of the
current MC generators~\cite{Buckley:2011ms}, and the CGC factorization
encompassing the physics of high gluon density to LLA. These two
formalisms have very different mathematical structures, so it is hard to
see how to merge them within a same event generator: collinear
factorization is based on the notion of (integrated) parton distribution
function and its DGLAP evolution, while the CGC formalism uses the
language of classical color fields distributed with a functional weight
function obeying JIMWLK. The Lund Dipole Cascade Model~\cite{Gustafson}
provides a meaningful interpolation between the two, but it cannot
accurately describe the high--$Q^2$ regime which is so important for jet
physics or BSM searches at the LHC. One can also envisage event
generators based on the CGC approach~\cite{Gelis:2010nm}, but once again
they should not be accurate enough at high $Q^2$ (and rather cumbersome
in practice).

\begin{figure}[htb!]
\begin{center}
\includegraphics[width=0.95\textwidth]{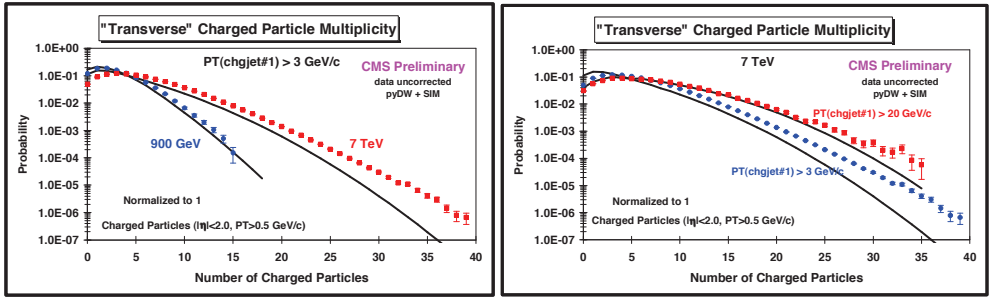}
\end{center}
\caption{\label{fig:Field}\sl CMS preliminary data (dots) for particle
production in p+p collisions vs. the respective predictions of
the Pythia DW tune (the continuous lines) \cite{Field}. Left: 2 different
center-of-mass energies. Right: two different transverse resolutions
(as determined by the transverse momentum of the leading jet).}
\end{figure}

In view of this, it is important to keep developing the conventional MC
event generators (Ariadne, Herwig, Hijing, Pythia, Sherpa etc
\cite{Buckley:2011ms,Deng:2010mv}), not only by successive `re-tunings'
of the free parameters, but also by (at least semi--heuristically)
improving the treatment of the semi--hard sector, using guidance from the
recent theoretical progress. The need for such an improvement is also
demonstrated by the difficulty of the present generators to reproduce the
soft and semi--hard parts of the underlying and minimum-bias events at
the LHC, as reviewed at ISMD2010 by R.~Field~\cite{Field}. To quote Rick,
``Pythia Tune DW describes the LHC distributions fairly well, but not
perfectly''. More precisely, this particular Pythia tune which was
calibrated to fit the UE data at CDF at 1.96~TeV is systematically
underestimating the hadronic activity (in terms of either multiplicity or
transverse energy) in p+p collisions at the LHC for the underlying events
accompanying a leading jet (CMS) or a leading particle (ATLAS) with
transverse momentum $k_\perp^{\rm jet}$ of a few GeV (say, below 10~GeV).
The discrepancies become higher with increasing energy at a given
$k_\perp^{\rm jet}$ (from 900~GeV to 7~TeV) and for decreasing
$k_\perp^{\rm jet}$ at a given energy (see Fig.~\ref{fig:Field}).
Moreover, they persist (although to a lower degree) after re-tuning the
MC on the basis on the CMS data (`Pythia Tune Z1')~\cite{Field}. This
shows that, beyond merely re-tuning, we need a better description (like
CGC--inspired models) for multiparticle interactions and the partonic
distributions at small $x$ and semi--hard $k_\perp$.

Another traditional topics of ISMD, which has been also present at the
2010 edition, concerns the attempts to model the low $k_\perp$ part of
the multiparticle production. Some of these models are merely fits with a
heuristic physical interpretation. For instance,
A.~Rostovtsev~\cite{Rostovtsev} has shown that a combination (somewhat
reminiscent of the photon spectrum in the Solar flares) of a
thermal--like exponential and an inverse power law  can describe the
low--$k_\perp$ part of the spectrum within a wide range of energies and
for various processes (p+p, $\gamma+\gamma$, DIS, A+A).
G.~Wilk~\cite{Wilk} emphasized that a power--law spectrum is not
necessarily a signal of perturbative--like, hard QCD, behaviour: a power
distribution of the Tsallis type can be also generated via fluctuations
in the (local) temperature associated with a thermal distribution.
Moreover fluctuations in temperature ($T$) are equivalent to fluctuations
in the interaction volume ($V$) provided the total energy $E$ is kept
constant, since $E\sim V T^4$. W.~Ochs~\cite{Ochs} has argued that the
zero momentum limit of the spectra should be independent of the energy
since dominated by the bremsstrahlung of soft gluons which cannot resolve
the partonic structure of the final state, but only the overall color
charge of the primary sources. V.~Abramovsky~\cite{Abramovsky} used a
model combining valence quarks and string hadronization to argue that
particle production should be different in p+p vs. p+$\bar{\rm p}$
collisions, a difference that could be observed by comparing
high--multiplicity events at the LHC and the SPS (at the common energy
$\sqrt{s}=900$~GeV). S.~Todorova~\cite{Todorova} has described a
refinement of the Lund string model in terms of a helix string which can
describe the `bump' observed in the LEP (DELPHI) data at $k_\perp\sim
0.5$~GeV. H.~Gr\"onqvist~\cite{Gronqvist} proposed to measure the elastic
p+p cross--section at the LHC by tagging the forward bremsstrahlung
photons with the Zero Degree Calorimeter at the CMS. Such a measurement
could also be used to check the relative alignement of the ZDCs and of
the Roman Pot detectors.

Besides the single particle spectra, the {\em correlations} in the
multiparticle production at soft momenta represent a challenge for the
theorists and the topics of one of the traditional sessions of ISMD. In
his introductory talk to this session, A.~Bialas~\cite{Bialas} made some
remarks on the possible origin of four important classes of correlations:
\texttt{(i)} the negative binomial distribution in the multiplicity,
which reflects the particle production by a superposition of various
sources, possibly grouped into `clans'; \texttt{(ii)} the
forward--backward correlations in rapidity, which give information about
the early stages of the collision; \texttt{(iii)} the balance functions
for pairs of charged particles, as measured by
STAR~\cite{Aggarwal:2010ya}, which are narrow in rapidity (and narrower
in central A+A collisions as compared to peripheral ones, or to p+p),
thus suggesting that charges are created in the late stages of the
collisions, just before the freeze-out, and \texttt{(iv)} the HBT
correlations, for which the descriptions based on the Lund string picture
\cite{Csorgo:2004sr} and respectively on intermittency \cite{iancuBialas}
can be related to each other by assuming a fluctuating string tension.
T.~Csorgo~\cite{Csorgo} proposed the measurement of the mass of the
$\eta'$ meson as a probe of chiral symmetry restoration in A+A
collisions. The expected reduction of this mass in the medium should lead
to an increase of the abundance of $\eta'$ in the final state, which in
turn can be measured using Bose--Einstein correlations. The respective
analysis of the RHIC data suggests indeed an in--medium reduction of the
$\eta'$ mass by at least 200~MeV~\cite{Csorgo}.

\section{Conclusions}

The 2010 edition has marked the entrance of the ISMD series of
conferences in the LHC era. It has demonstrated the vitality of the
original theme of this meeting --- the physics of multiparticle
interactions --- which nowadays lies at the heart of the QCD physics at
the LHC. It has also demonstrated the capacity of this meeting to
permanently renew itself by integrating new themes --- notably, the hard
and semi--hard partonic interactions ---, thus following and stimulating
the progress on both theoretical and experimental sides. All these themes
have met with important developments over the recent years, that I tried
to succinctly summarize here, with emphasis on the contributions
presented at ISMD2010. All these themes need further progress in order to
match the exigences of the present day high--energy experiments. The ISMD
series offers a privileged, almost unique, opportunity for fruitful
exchanges and cross-fertilizations between these various themes. Forty
years after, the International Symposium on Multiparticle Dynamics is
more than ever at the center of the scientific debate. {\sf ISMD, Happy
Anniversary !}

\section*{Acknowledgments}

I would like to thank the organizers of ISMD2010, especially Pierre van
Mechelen and Nick Van Remortel, for their efforts towards a most
instructive and enjoyable meeting in Antwerp and for offering me the
prestigious but challenging task to present this theory summary. I am
grateful to Hanna Gr\"onqvist, Fran\c cois Gelis, Cyrille Marquet and
Dionysis Triantafyllopoulos for reading the manuscript, useful comments
and related discussions.

\begin{footnotesize}

\end{footnotesize}

% ****************************************************************************
% END OF BIBLIOGRAPHY AREA
% ****************************************************************************

\end{document}